\documentclass[prd,twocolumn,nofootinbib, superscriptaddress]{revtex4-1}
\usepackage{epsfig}
\usepackage{graphicx}
\usepackage{palatino}
\usepackage[english]{babel}
\usepackage{hyphenat}
\usepackage{amsmath}
\usepackage{amssymb}
\usepackage{mathtools}
\usepackage{mathrsfs}
\usepackage{slashed}
\usepackage{epstopdf}
\usepackage{xcolor}
\usepackage{booktabs}
\definecolor{lcolor}{rgb}{0.,0.0,0.}
\definecolor{citcolor}{rgb}{0,0.,0.5}
\usepackage[breaklinks,colorlinks,urlcolor=blue,citecolor=blue,linkcolor=blue]{hyperref}
\usepackage{multirow}
\usepackage{ltablex}
\newcommand{\beq}{\begin{eqnarray}}
\newcommand{\eeq}{\end{eqnarray}}
\newcommand{\dis}{\displaystyle}
\def\dd{{\rm d}}

\newcommand{\bem}{\begin{multline}}
\newcommand{\eem}{\end{multline}}
\newcommand{\beg}{\begin{gather}}
\newcommand{\eeg}{\end{gather}}

\newcommand{\nn}{\nonumber\\}

\newcommand{\ben}{\begin{eqnarray*}}
\newcommand{\een}{\end{eqnarray*}}

\newcommand{\secn}[1]{Section~1}
\newcommand{\appn}[1]{Appendix~1}

\long\def\comment#1{ }

\def\and{\quad\text{and}\quad}

\newcommand{\rmd}{{\rm d}}

\newcommand{\rme}{{\rm e}}

\newcommand{\abar}{\bar{\alpha}}

\begin{document}

\title{Dynamically groomed jet radius in heavy-ion collisions}

\author{Paul Caucal}
\email {pcaucal@bnl.gov}
\affiliation{Physics Department, Brookhaven National Laboratory, Upton, NY 11973, USA}
\author{Alba Soto-Ontoso}
\email {alba.soto@ipht.fr}
\affiliation{Universit\'e Paris-Saclay, CNRS, CEA, Institut de physique th\'eorique, 91191, Gif-sur-Yvette, France}
\author{Adam Takacs}
\email{adam.takacs@uib.no}
\affiliation{Department of Physics and Technology, University of Bergen, Bergen 5020, Norway}

\begin{abstract}
We explore the ability of a recently proposed jet substructure technique, Dynamical Grooming, to pin down the properties of the Quark-Gluon Plasma formed in ultra-relativistic heavy-ion collisions. In particular, we compute, both analytically and via Monte-Carlo simulations, the opening angle $\theta_g$ of the \textit{hardest} splitting in the jet as defined by Dynamical Grooming. Our calculation, grounded in perturbative QCD, accounts for the factorization in time between vacuum-like and medium-induced processes in the double logarithmic approximation. We observe that the dominating scale in the $\theta_g$-distribution is the decoherence angle $\theta_c$ which characterises the resolution power of the medium to propagating color probes. This feature also persists in strong coupling models for jet quenching. We further propose for potential experimental measurements a suitable combination of the Dynamical Grooming condition and the jet radius that leads to a pQCD dominated observable with a very small sensitivity ($\leq 10\%$) to medium response. 
\end{abstract}
\maketitle
\section{Jet substructure in heavy-ion collisions}
\label{sec:intro}
The use of jet substructure techniques in heavy-ion collisions is ramping up, see Refs.~\cite{Andrews:2018jcm,Cunqueiro:2021wls} and references therein. From a theoretical viewpoint, there are certain advantages when considering observables defined in terms of one or a few jet constituents with respect to global ones such as fragmentation functions~\cite{CMS:2014jjt,ATLAS:2017nre,Caucal:2020xad} or jet shapes~\cite{CMS:2013lhm,Acharya:2018uvf,KunnawalkamElayavalli:2017hxo,Chien:2015hda}. In particular, jet substructure observables can be engineered to enhance the sensitivity to certain regions of the radiation phase-space where perturbative QCD effects dominate, thus enabling first principles calculations. Experimentally, fully corrected jet substructure measurements are now available in heavy-ion collisions both at RHIC and LHC energies~\cite{STAR:2021kjt,ALICE:2021obz}. They are highly complementary to the rich dataset recorded in $pp$ collisions both for low~\cite{STAR:2020ejj,ALICE:2021njq} and high-$p_t$ jets, e.g.~\cite{CMS:2018ypj,ATLAS:2019mgf}. 
 
Up to now, the jet substructure program in the heavy-ion community has strongly focused on SoftDrop (SD) observables~\cite{Dasgupta:2013ihk, Larkoski:2014wba}. They are defined in terms of the kinematics of the first branching in an angular-ordered splitting tree whose momentum sharing\footnote{For a given splitting with prongs $i$ and $j$, the momentum sharing is defined as $z={\rm{min}}(E_i+E_j)/(E_i+E_j)$, where $E$ represents the energy.} $z$ satisfies the so-called SoftDrop condition, $z>z_{\rm cut}\theta^\beta$, where $\theta$ is the relative angle of the branching and ($z_{\rm cut},\beta$) are free parameters. One of such observables is the distribution of $z$ values that pass the SD cut. In vacuum, the $z_g$-distribution for $\beta\!=\!0$ is known to scale as the Dokshitzer-Gribov-Lipatov-Altarelli-Parisi (DGLAP) splitting function \cite{gribov1972deep,altarelli1977asymptotic,Dokshitzer:1977sg} to lowest order in pQCD, i.e. $\dd\sigma/\dd z_{g}\sim 1/z_g$~\cite{Larkoski:2015lea,Cal:2021fla}. In the medium, several ingredients are expected to play a role. On the one hand, assuming that the interaction between the high energetic jet and the medium is dominated by multiple, soft scatterings, an enhancement of low-$z_g$ splittings is expected due to the $\propto\!z^{-3/2}$ scaling of the medium-induced radiative spectrum~\cite{Baier:1996sk,Zakharov:1997uu}. On the other hand, incoherent energy loss leads to more asymmetric splittings being suppressed with respect to the vacuum baseline~\cite{Casalderrey-Solana:2012evi}. These two competing effects are, in general, hard to disentangle and their relative magnitude will depend on the jet $p_t$ together with the parameters of the grooming condition, see Ref.~\cite{Caucal:2019uvr}. The first $z_g$ measurement in heavy-ion collisions by CMS showed a steeper $z_g$ distribution with respect to the vacuum baseline~\cite{Sirunyan:2017bsd}. The theoretical interpretation of this softening remains unclear given that the data has been quantitatively reproduced by models whose in-medium dynamics are disparate, e.g. Refs.~\cite{Mehtar-Tani:2016aco,Chien:2016led} related the enhancement of soft particles to the medium-modified splitting functions while Ref.~\cite{Milhano:2017nzm} proposed a medium-response based description of the data. In addition, no obvious modification of the $z_g$ distribution has been observed at RHIC energies~\cite{STAR:2020ejj}. The simultaneous description of both data sets has been provided in Ref.~\cite{Chang:2017gkt} where it was argued that the energy-dependence of the $z_g$ distribution was dominated by the density of the QGP together with coherent energy loss. Nevertheless, a back of the envelope calculation shows that the $k_t$ of the splitting probed by the RHIC measurement can be as large as $k_t\propto z p_t R$, that is $k_{t}\!=\!1$~GeV for $p_t\!=\!25$~GeV, $R\!=\!0.4$ and $z\!=\!0.1$. Thus, non-perturbative dynamics are expected to play a role and a purely pQCD approach might not be well suited. Along these lines, the impact of the fluctuating thermal background on the $z_g$ distribution, among other observables, has been recently assessed in Refs.~\cite{Andrews:2018jcm,Mulligan:2020tim}. The authors showed that mistagged splittings can induce a non-negligible contribution ($\mathcal{O}(10\%)$) that mimic a jet quenching signal. This fact lead ALICE to increase the value of $z_{\rm cut}$ from the standard value in $pp$, i.e. $z_{\rm cut}=0.1$, to 0.2 in their recent publication~\cite{ALICE:2021obz}. Also, in this recent measurement, the $z_g$ is integrated over all possible angles of the splitting. These two combined ingredients, i.e. the enhanced $z_{\rm cut}$ and the integration over the angles, result into an unmodified $z_g$-distribution.

Another SoftDrop observable that has been studied in the context of jet quenching is the opening angle of the SD splitting, $\theta_g$. The physics motivation in this case is related to the intrinsic medium angular scale $\theta_c$ that divides the radiation phase-space into resolved and unresolved splittings~\cite{Mehtar-Tani:2010ebp,Casalderrey-Solana:2011ule,Mehtar-Tani:2014yea}. In a nutshell, splittings with $\theta > \theta_c$ lose more energy than those with $\theta < \theta_c$. Then, the steeply falling nature of the jet $p_t$ spectrum leads to a filtering effect such that only quasi-collinear splittings pass the selection cut and thus a narrowing of the $\theta_g$ distribution when compared to $pp$ is to be expected~\cite{Rajagopal:2016uip}. There is a competing effect that leads to a broadening of the $\theta_g$ distribution, namely transverse momentum diffusion of each of the resolved branches when $\theta_g > \theta_c$~\cite{Ringer:2019rfk}. Recent measurements by ALICE~\cite{ALICE:2021obz} indicate an overall narrowing of the $\theta_g$ distribution with respect to $pp$. The physics mechanism driving this observation is far from being settled given that models with~\cite{Caucal:2018dla} and without a color (de)coherence mechanism
~\cite{Casalderrey-Solana:2019ubu,Putschke:2019yrg,Ringer:2019rfk} are able to semi-quantitatively describe the data. 

Overall, Ref.~\cite{ALICE:2021obz} showed that the only models that correctly reproduce both $z_g$ and $\theta_g$ data are (i) JetMed~\cite{Caucal:2018dla,Caucal:2019uvr}, where the coherence angle is built-in, (ii) the Hybrid~\cite{Casalderrey-Solana:2019ubu} with a fully incoherent energy loss picture and (iii) the JetScape~\cite{Putschke:2019yrg} result using MATTER+LBT~\cite{Majumder:2013re,He:2015pra}, a model completely agnostic to $\theta_c$. Given the lack of consensus in the theoretical interpretation of the SoftDrop measurements, a natural question is whether instead of merely adopting jet substructure techniques that were designed by the $pp$ community, one should develop specific tools best suited to in-medium jet physics. As we have already mentioned, the necessity to double the value of $z_{\rm cut}$ to mitigate the impact of the underlying event highlights the specificities of heavy-ion collisions. Another example along this direction of thought is Ref.~\cite{Apolinario:2020uvt} where a jet clustering algorithm that uses as metric the formation time of the splitting is explored. Regarding groomers, Ref.~\cite{Mehtar-Tani:2019rrk} proposed the Dynamical Grooming procedure which relies on identifying the hardest splitting in the QCD shower as defined by the maximal value of the so-called `hardness' variable 
\beq
\label{eq:hardness}
\kappa^{(a)}=\frac{1}{p_{t,{\rm jet}}}z(1-z)p_t\left(\frac{\theta}{R}\right)^{a}\, ,
\eeq
where $a\!>\!0$ is a continuous free parameter and $(p_{t,{\rm jet}},R)$ the transverse momentum and cone size of the jet. The most natural values of $a$ from a heavy-ion perspective are $a\!=\!1,2$. Indeed, when setting $a=1$ the splitting with the largest transverse momentum, $k_t$, is selected. This choice is interesting from the point of view of probing the quasi-particle nature of the QGP. In short, rare, hard scatterings between the propagating color probe and the medium lead to a $\propto 1/k_t^4$ scaling of the $k_t$-distribution's tail. Thus, an enhancement in the distribution at large-$k_t$ could serve as the smoking gun for Rutherford-like scatterings~\cite{DEramo:2018eoy,Caucal:2021lgf}. Further, selecting $a\!=\!2$ corresponds to the splitting with the shortest formation time. In principle, the shorter the formation time of a splitting is, the larger in-medium modifications with respect to the vacuum dynamics are expected. One could then subdivide a sample of jets into short and large formation time splittings and compare the size of the modifications with respect to the $pp$ result~\cite{Apolinario:2020uvt}.
  
So far, the Dynamical Grooming technique has been only applied to $pp$ physics both theoretically and experimentally. In particular, Ref.~\cite{Caucal:2021bae} presented a thorough examination of the analytic structure of dynamically groomed observables that lead to a quantitative description of the ALICE preliminary data~\cite{Mulligan:2020cnp, Ehlers:2020piz}. Equipped with a solid understanding of the vacuum benchmark, we extend the theoretical calculation to in-medium jet physics. In this paper, we focus on the angle of the splitting tagged by dynamical grooming, while the relative $k_t$ will be presented in a separate publication~\cite{kt:paper}. The goal of this paper is to showcase the main physics ingredients that enter into the theoretical calculation of $\theta_g$ and ease the interpretation of Monte-Carlo results. In particular, we demonstrate that this observable is highly sensitive to the Quark-Gluon plasma resolution angle. 

The analytic calculation is presented in Section.~\ref{sec:theory}. After a brief reminder of the vacuum calculation, we move on to the in-medium theoretical analysis in Section~\ref{sec:medium}. We build up our toy in-medium shower incrementally such that the impact of each ingredient in the $\theta_g$ distribution can be clearly disentangled. The final theoretical curves can be found in Section~\ref{sec:final-theory} where we make use of the Kolmogorov-Smirnov metric to quantify the discriminating power of the observable. Then, we compare our analytic estimates to a pQCD based Monte-Carlo in Section~\ref{sec:jetmed}. We show that the effects that we observe are qualitatively robust even in a strong coupling description of the medium by using the Hybrid model. Finally, in Section~\ref{sec:medium-response}, we evaluate the impact of the medium response using state-of-the-art jet quenching Monte-Carlo generators and present a systematic study of the best setup to enhance the impact of pQCD physics on this observable experimentally. We conclude and outline some additional ideas in Section~\ref{sec:conclusions}. The numerical routines used in this publication can be found in~\cite{python_git}.

\section{Theoretical setup}
\label{sec:theory}
We begin by formulating the Dynamical Grooming technique in its most general terms, i.e. independently of whether the emission takes place in vacuum or in the medium. Here, we provide the main formulas and refer the reader to Refs.~\cite{Mehtar-Tani:2019rrk,Caucal:2021bae} for a more detailed discussion on their derivation. Our main assumption is that we work in the soft and collinear limit such that $z\!,\theta\!\ll \!1$ and we can neglect momentum degradation along the jet primary branch. Then, Eq.~\eqref{eq:hardness} reduces to
\beq
\label{eq:hardness-dla}
\kappa^{(a)}=z\left(\frac{\theta}{R}\right)^a\, .
\eeq
Next, we take the $\kappa\ll 1$ limit in order for resummation techniques to apply \cite{Caucal:2021bae}. The probability distribution for a splitting to be the hardest in a QCD jet can be written as 
\beq
\label{eq:prob-dist}
\dis\frac{\dd^2 \mathcal P(z,\theta|a)}{\dd z\dd \theta}=\frac{\dd^2\widetilde{P}(z,\theta)}{\dd z\dd \theta}\Delta(\kappa|a)\,,
\eeq
where $\dd^2\widetilde{P}(z,\theta)$ is a branching kernel that represents the probability of a splitting with ($z,\theta$) to take place along the jet fragmentation and $\Delta(\kappa|a)$ is a Sudakov form factor that is the probability of no emission with hardness larger than $\kappa^{(a)}$. These two functions are related by
\begin{align}
\label{eq:sudakov_general}
\ln \Delta(\kappa|a) &= -\displaystyle\int_0^{1} \dd z' \displaystyle\int_0^R \dd\theta'\,\frac{\dd^2\widetilde{P}(z',\theta')}{\dd z'\dd \theta'} \nn
&\times\Theta\left(z'\left(\theta'/R\right)^a-\kappa^{(a)}\right)\,.
\end{align}
Although left implicit, note that the branching kernel and, thereby, the Sudakov form factor appearing in Eqs.~\eqref{eq:prob-dist} and \eqref{eq:sudakov_general} depend on the color representation of the jet initiating parton. In addition, to guarantee the collinear safety of the Sudakov form factor, we require $a\!>\!0$.

In this work, we are interested in the angular distribution of the splitting tagged by Dynamical Grooming. It is obtained directly from Eq.~\eqref{eq:prob-dist} by marginalising over $z$. That is, 
\begin{align}
\label{eq:thetag-general}
\left.\frac{1}{\sigma} \frac{\dd\sigma}{\dd \theta_g} \right|_a &= \int_0^1 \dd z \,  \frac{\dd^2\mathcal P(z,\theta|a)}{\dd z\dd\theta}\delta(\theta-\theta_g)\, .
\end{align}
Note that this differential distribution is self-normalized by definition.  

The purpose of the next sections is to compute Eq.~\eqref{eq:thetag-general} for vacuum and in-medium jets.
\subsection{Vacuum recap}
\begin{figure}
     \includegraphics[scale=0.8]{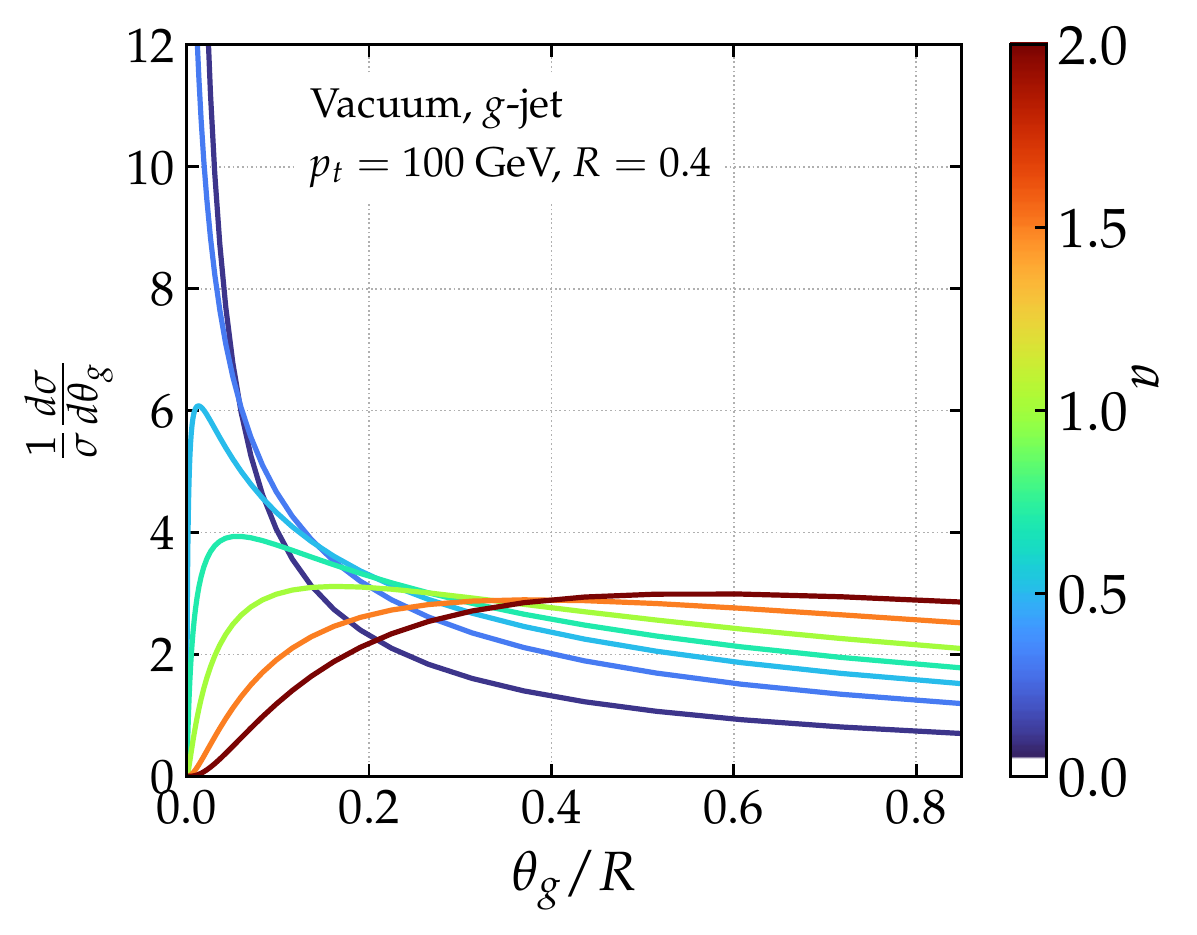}
    \caption{The $\theta_g$-distribution in vacuum for various values of the Dynamical grooming parameter $a$. The peak position of the distribution scales as $1/a$.}
    \label{fig:vacuum-thetag}
\end{figure}

In the Double Logarithmic Approximation (DLA) on which we rely throughout this paper\footnote{See Ref.~\cite{Caucal:2021bae} for a higher order computation.} it is sufficient to consider the branching kernel in the soft ($z\!\ll1\!$) and collinear ($\theta\!\ll\!1$) limit, 
\beq
\label{eq:branch-dla}
\frac{\dd^2\widetilde{P}^{\rm vac}(z,\theta)}{\dd z\dd \theta}=\dis\frac{2\alpha_sC_R}{\pi} \displaystyle\frac{1}{z\theta}\,,
\eeq
where $C_R$ is the Casimir factor of the representation of the leading parton. At this level of accuracy, the strong couplingg constant is fixed to the hardest transverse momentum scale of the problem $Q=p_{t,{\rm jet}}R$, namely $\alpha_s\equiv\alpha_s(Q)$. The remaining integrations in Eq.~\eqref{eq:thetag-general} can be carried out analytically and yield
\begin{align}
\label{eq:thetag_dla}
\dis\frac{1}{\sigma}\dis\frac{\dd\sigma}{\dd \theta_g} = \frac{1}{\theta_g}\dis\sqrt{\bar{\alpha}\pi a}\left[{\rm{erf}}
\left(\sqrt{\bar{\alpha}a}\ln\left(\frac{\theta_g}{R}\right)\right)+1\right]\,,
\end{align}
with $\bar\alpha\equiv\alpha_sC_R/\pi$. This distribution is shown in Figure~\ref{fig:vacuum-thetag} as a function of the grooming parameter $a$. We observe that the lower the value of $a$, the more collinear the `hardest' (or tagged) splitting is. In fact, this is confirmed analytically after taking the first derivative of Eq.~\eqref{eq:thetag_dla}, we obtain the maximum of the distribution to be
\begin{equation}
\label{eq:theta-cut}
\ln\left(\frac{1}{\theta_{\rm max}}\right)=\frac{1}{2a\abar}+\mathcal{O}(1)\,.
\end{equation}
To conclude this vacuum recap, we would like to emphasise that the regions of phase-space that the $\theta_g$ observable explores are heavily correlated with the choice of $a$. In particular, setting $a\geq 1$ leads to the observable being sensitive to wide angle dynamics. This observation will play an important role in the next section, where we extend the calculation 
to account for in-medium jet evolution. 
\subsection{In-medium calculation}
\label{sec:medium}
\begin{figure}
     \includegraphics[width=\columnwidth]{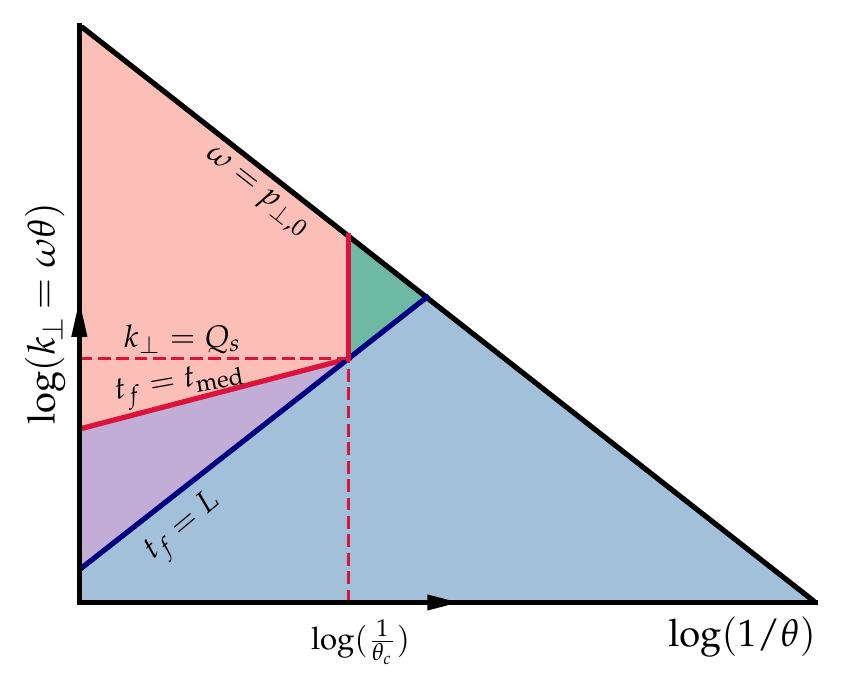}
    \caption{Schematic Lund plane representation of the relevant regions of phase-space for an in-medium jet.
    The physical meaning of the different lines is explained in the main text.}
    \label{fig:medium-lp}
\end{figure}
After more than two decades of active theoretical work, our understanding on how to describe the fragmentation process of a highly-energetic parton in the medium has significantly improved. One of the building blocks of an in-medium parton shower is the medium-induced radiative spectrum. Recently, it has been computed beyond its two asymptotic limits: (i) rare, hard~\cite{Gyulassy:2000fs} and (ii) multiple, soft scatterings~\cite{Baier:1996sk, Zakharov:1997uu} between the jet and the medium constituents. A few numerical approaches can be found in Refs.~\cite{Caron-Huot:2010qjx,Feal:2018sml,Andres:2020vxs,Andres:2020kfg}, while a novel expansion scheme has been proposed at the analytic level~\cite{Barata:2021wuf} and applied to some global observables in Refs.~\cite{Mehtar-Tani:2021fud,Takacs:2021bpv}. Significant advances have been made on the evolution of this gluonic cascade~\cite{Barata:2021byj,Adhya:2021kws,Blanco:2021usa,Rohrmoser:2021yrh} and its cross-talk with the vacuum evolution~\cite{Caucal:2018dla}. In addition, semi-analytic approaches~\cite{Andres:2019eus,Mehtar-Tani:2021fud} are now incorporating realistic collision geometries and local medium properties from the hydrodynamic evolution of the medium.

In this work, we make three major simplifications to facilitate analytic manipulations. First, we restrict ourselves to the double logarithmic limit of pQCD. In this limit, there is a factorization in time between vacuum-like emissions (enhanced by large soft and collinear logarithms) and medium-induced emissions \cite{Caucal:2018dla}. Further,  we treat the medium as a static, brick of length $L$. This allows us to neglect the time dependence of the quenching parameter, i.e. $\hat q(t)\equiv \hat q \Theta(L-t)$. In the first part of this Section, $L$ is taken as a constant, while by the end of it we evaluate the impact of its fluctuations by using a probability distribution that mimics the collision geometry. Lastly, we describe the jet-medium interaction in the multiple soft, scattering approximation. That is, we only account for inelastic collisions with low momentum exchanges between the propagating parton and the medium constituents. These interactions lead to medium-induced emissions together with a Gaussian diffusion in transverse space characterised by the momentum scale $Q^2_s\equiv \hat q L$.

Within this simplified scenario, the phase-space for the first branching can be sketched in a Lund-plane representation~\cite{Dreyer:2018nbf} as the one provided in Fig.~\ref{fig:medium-lp}, where we use the transverse momentum $k_\perp$ of the emission and its opening angle $\theta$ as coordinates. We can approximate the quantum mechanical formation time of the emission as $t_f\simeq 2/(k_\perp\theta)=2/(\omega\theta^2)$. Let us discuss the different regions in Fig.~\ref{fig:medium-lp}:

\begin{itemize}
 \item{\textit{Blue region:}} The most obvious constraint on the radiative phase space is generated by the finite length of the medium: emissions with $t_f>L$ are created outside of it and thus their fragmentation process develops as in vacuum. 
\item{\textit{Red and purple regions:} On the other hand, branchings with $t_f<L$ can be classified into two categories: vacuum-like (VLEs) and medium-induced (MIEs). In this case, the relevant scale arises by considering that, any emission inside the medium have a minimum transverse momentum set by the one acquired via  multiple soft collisions during its formation time:
\beq 
k^2_\perp\ge k^2_{\perp,{\rm med}}\equiv \hat qt_f\, .
\eeq 
Emissions which saturate this constraint, $k_\perp=k_{\perp,{\rm med}}$, are medium-induced, while emissions with $k_\perp^2\gg k_{\perp,{\rm med}}$ are vacuum-like\footnote{Large $k_\perp$ emissions can also be triggered by single hard collisions with a medium scattering center, but we neglect this kind of contribution in this study.}. In terms of formation time, the latter condition becomes $t_f\ll t^{\rm med}_f$ with the formation time of a medium-induced emission given by
\beq 
t^{\rm med}_f = \sqrt{2\omega/\hat q}\, ,
\eeq 
Therefore, vacuum-like emissions have much shorter formation times that medium-induced ones. Consequently, vacuum-like emissions are vetoed in the $t^{\rm med}_f < t_f \ll L$ region, i.e. the purple area in Fig.~\ref{fig:medium-lp}~\cite{Caucal:2018dla}. Besides $t^{\rm med}_f$, there is yet one more scale that plays a prominent role in this paper, i.e. the decoherence angle $\theta_c$ given by (see e.g. Ref.~\cite{Casalderrey-Solana:2011ule,Casalderrey-Solana:2012evi})
\beq
\theta_c=\frac{2}{\sqrt{\hat q L^3}}\, .
\eeq 
As we have already mentioned, this angular scale separates resolved from unresolved emissions. The purple region contains branchings with $\theta>\theta_c$. As such, the two prongs act as independent emitters after the splitting.
}

\item{\textit{Green region:} In this area, splittings are typically vacuum-like, but are never resolved by the medium given that $\theta<\theta_c$. These emissions lose energy as a single color charge.}
\end{itemize}

In what follows, we present analytic estimates for the dynamically groomed $\theta_g$-distributions of splittings generated in the regions of phase space that we have discussed above. We would like to remark that it is not the aim of this paper to provide precise analytic predictions, but rather to illustrate the main physics ingredients that enter into the theoretical calculation of $\theta_g$ in order to facilitate an interpretation of the Monte-Carlo results that will be shown by the end of this manuscript. 
\subsubsection{Vacuum-like emissions}
Formally, the only leading-logarithmic effect of the medium on the dynamically groomed distributions is caused by the veto constraint on vacuum-like emissions in the presence of a dense medium (see purple region in Fig.~\eqref{fig:medium-lp}). 
As we have described above, emissions whose formation time satisfy $t^{\rm med}_f<t_f<L$ and whose angle is $\theta>\theta_c$ are vetoed, and therefore, Eq.~\eqref{eq:prob-dist} is amended accordingly:
\begin{equation}\label{eq:p-veto}
\frac{\dd^2 \mathcal P^{\rm vle}}{\dd z\dd\theta}=\frac{\dd^2\widetilde{P}^{\rm vac}(z,\theta)}{\dd z\dd\theta}\Theta_{\notin \textrm{veto}}(z,\theta)
\Delta_{\notin\rm veto}(\kappa|a)\, ,
\end{equation}
where $\dd^2\widetilde{P}^{\rm vac}(z,\theta)$ is given by Eq.~\eqref{eq:branch-dla}. In this case, the Sudakov form factor reads
\begin{align}\label{eq:suda-veto}
\ln\Delta_{\notin\rm veto}(\kappa|a) &= -\displaystyle\int_0^1 \dd z' \displaystyle\int_0^R \dd\theta'\,\frac{\dd^2\widetilde{P}^{\rm vac}(z',\theta')}{\dd z'\dd\theta'} \nn
&\times\Theta_{\notin \textrm{veto}}(z',\theta')\ \Theta\left(z'(\theta'/R)^a-\kappa\right)\, ,
\end{align}
with
\begin{align}
\label{eq:veto-cond}
 \Theta_{\notin \textrm{veto}}(z,\theta)&=1-\Theta(\theta-\theta_c)\Theta(t_f-t^{\rm med}_f)\Theta(L-t_f)\nn
                                        &=1-\Theta(\theta-\theta_c) \Theta(2\hat{q}-z^3p_t^3\theta^4) \nn
                                        &\times\Theta(zp_t\theta^2L-2)\,.
\end{align}
Note that these medium boundaries are known at double logarithmic accuracy only, meaning that the numerical pre-factors (such as the factors $2$ in the veto constraint) are not under control and have been chosen in this way for convenience. Consequently, one can perfectly neglect single logarithmic terms such as hard collinear or running couplingg corrections in Eq.~\eqref{eq:p-veto} since our calculation cannot be more accurate than double-log due to medium-related uncertainties in the phase space for vacuum-like emissions.

\begin{figure}
     \includegraphics[width=\columnwidth]{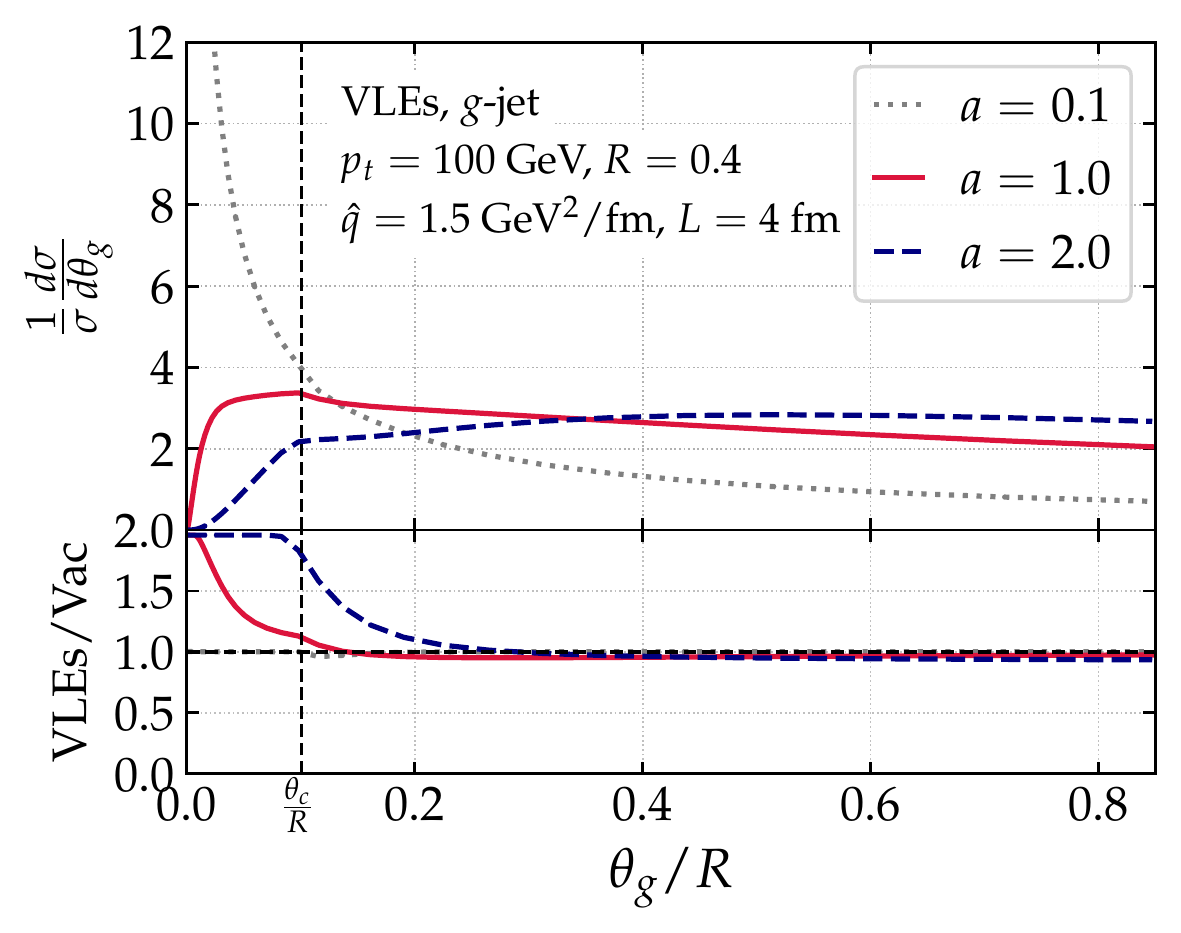}
    \caption{The $\theta_g$-distribution for gluon jets in medium including only vacuum-like emissions in the phase space of Fig.~\ref{fig:medium-lp}, for various values of the Dynamical Grooming parameter $a$. The lower panel shows the ratio to the pure vacuum expectation presented in Fig.~\ref{fig:vacuum-thetag}. The veto region forbids some of the wide angle emissions and thus enhances emissions with small $\theta_g$.}
    \label{fig:vac-veto-thetag}
\end{figure}

The calculation of Eq.~\eqref{eq:suda-veto} is provided in Appendix~\ref{app:sudveto}. The integral over $z'$ is done analytically, while the remaining integral over $\theta'$ is performed numerically to avoid the difficulties related to the complicated shape of the integration domain. In Figure~\ref{fig:vac-veto-thetag} we present the impact of the veto constraint on the $\theta_g$ distribution. The kinematic parameters are chosen to resemble an ALICE-like setup~\footnote{Note that the factorised picture described in Fig.~\ref{fig:medium-lp} is best suited for large $p_t$ jets and, therefore, the ALICE-like kinematics is not optimal.}: $\alpha_s\!=\!0.2, R\!=\!0.4, p_t\!=\!100$~GeV/c, $L = 4$~fm and $\hat q = 1.5$~GeV$^2$/fm. The medium parameters are tuned such that our final theoretical result agrees with the nuclear modification, $R_{AA}$, in the ALICE jet selection window~\cite{ALICE:2019qyj}. In addition, the jet $p_t$ always refers to the final transverse momentum, i.e. after quenching, although when energy loss is absent this value coincides with the $p_t$ of the initiator. We observe how the presence of the veto region leads to a relative narrowing of the distribution for $a\!=\!1,2$. This is expected given that the veto region mainly prohibits large angle emissions and thus, collinear radiation is enhanced. Due to the self-normalization of the observable, this leads to a depletion of wide angle splittings. In the case of $a\!=\!0.1$, the effect is negligible since it tags narrow splittings by construction. Overall, the effects are only sizeable for $\theta_g/R\ll1$. Our main interest in this paper is to design an observable that enhances the sensitivity to the critical angle $\theta_c$ and that's the reason why we choose an angular observable such as the $\theta_g$-distribution. If instead one would like to maximise the impact of the veto region, it would be more convenient to explore observables with large values of $a$, like the groomed mass ($m^2_g\sim z\theta^2$), such that the tagging condition is parallel to the $t_{\rm med}$ line in Fig.~\ref{fig:medium-lp}.

%
\subsubsection{Medium-induced emissions}
The dynamically tagged splitting can also be a medium induced emission. The differential probability for these type of emissions, within a multiple scattering description of the parton-medium interaction, is given by a convolution of the BDMPS-Z energy spectrum with the angular distribution $\mathcal{B}(z,\theta)$ produced via transverse momentum broadening, i.e.
\beq
\label{eq:Pbranch-mie}
	\dis\frac{\dd^2\widetilde{P}^{\rm mie}}{\dd z\dd\theta}
	\underset{\omega\ll\omega_c}{\underset{k_\perp \ll Q_s}{\approx}} \bar \alpha_{s,\rm med} \sqrt{\dis\frac{2\omega_c}{z^3 p_t}}\Theta(\omega_c-z p_t) \mathcal{B}(z, \theta)\,,
\eeq 
where $\omega_c=\hat q L^2/2$ is the maximum energy that an emission can acquire as it corresponds to $t_f^{\rm med}=L$. In principle, the emission's energy is also bounded from below by the Bethe-Heitler frequency $\omega_{\rm{BH}}\propto \mu^4/\hat q$, with $\mu$ an infrared regulator of the order of the Debye mass ($\sim1$~GeV). That said, these soft emissions are suppressed by the Sudakov form factor in DyG observables and, consequently, this infrared physics is irrelevant. In this formula, the strong couplingg constant $\alpha_{s,\rm med}$ should be evaluated at the typical transverse momentum scale of a medium induced emission, $k_{\perp,\rm med}$. However, following the vacuum calculation, we shall consider $\alpha_{s,\rm med}$ as a constant parameter to be fixed, just like the other free parameters, by comparing our analytic model to the jet $R_{AA}$ observable in a given $p_{t,{\rm jet}}$ window. Note that $\alpha_{s,\rm med}$ can be distinct from its vacuum counterpart $\alpha_s$.

The factorization of the exact spectrum into the product of the time-averaged broadening distribution and the energy spectrum is only valid in the soft $\omega\ll \omega_c$ and collinear limit $k_\perp^2\ll Q_s^2$, i.e. for short-formation time emissions compared to the medium size. Such emissions can happen anywhere along the jet path length. Therefore, $\mathcal{B}(z,\theta)$ describes the transverse diffusion of the emission and is given by the average over the emission time $t\in[0,L]$ of a Gaussian distribution in $k_\perp\simeq \omega \theta$ with variance $\hat{q}(L-t)$, i.e.
\begin{align}
 	\mathcal{B}(z,\theta) &= \dis\frac{1}{L}\dis\int_0^L\dd t \dis\frac{2\omega^2\theta}{\hat q (L-t)}e^{-\frac{\omega^2\theta^2}{\hat q (L-t)}}\nonumber\\
	 &=2\theta\dis\frac{z^2p^2_t}{\hat q L}\Gamma\left(0,
	\dis\frac{z^2 p^2_t\theta^2}{\hat q L}\right)
\end{align}
where $\Gamma(a,z)=\int_z^\infty\rmd t\,t^{a-1}\rme^{-t}$ is the incomplete Gamma function. 

Although not realistic from a physics point of view, let us consider a jet evolving via primary medium-induced emissions \textit{only} (without VLEs), distributed according to Eq.~\eqref{eq:Pbranch-mie}. Then, the probability distribution for a medium-induced splitting to be the hardest in the shower is given by
\begin{align}
\label{eq:prob-dist-medium}
	\frac{\dd^2 \mathcal P^{\rm mie}}{\dd z\dd\theta} = \frac{\dd^2\widetilde{P}^{\rm mie}(z,\theta)}{\dd z\dd\theta}\Delta^{{\rm mie}}(\kappa|a) 
\end{align}
with the in-medium Sudakov form factor related to the medium-induced branching kernel as in Eq.~\eqref{eq:sudakov_general}. A straightforward calculation gives (for $\kappa<\omega_c/p_T$)
\begin{align}
\label{eq:sudakov_medium}
&\ln \Delta^{{\rm mie}}(\kappa|a) = -\displaystyle\int_0^{1} \dd z' \displaystyle\int_0^R \dd\theta'\,\frac{\dd^2 P^{\rm mie}}{\dd z'\dd \theta'}\Theta\left(z'\left(\theta'/R\right)^a-\kappa\right) \nn
& =-\bar{\alpha}_{s,\rm med}\sqrt{\frac{2\omega_c}{p_T}}\int_\kappa^{\omega_c/p_T}\frac{{\rm d} z'}{z'^{3/2}}\nn
& \times\left[z'^2\chi\Gamma\left(0,z'^2\chi\right)-\left(\frac{\kappa}{z'}\right)^{2/a}z'^2\chi\Gamma\left(0,\left(\frac{\kappa}{z'}\right)^{2/a}z'^2\chi\right)\right.\nn
&\left. -\exp\left(-z'^2\chi\right)+\exp\left(-\left(\frac{\kappa}{z'}\right)^{2/a}z'^2\chi\right)\right]\,,
\end{align}
with $\chi=Q^2/Q_s^2$ and $Q^2=p^2_tR^2$. Note that with Eq.~\eqref{eq:prob-dist-medium}, the normalization of $\dd^2 \mathcal P^{\rm mie}$ is not guaranteed. Indeed, when taking the limit $\kappa\to 0$ in Eq.~\eqref{eq:sudakov_medium} the Sudakov does not vanish, as it is the case for vacuum emissions, but rather tends to a constant. This difference arises from the absence of a collinear singularity in the medium-induced case. Therefore, in order to maintain the probabilistic interpretation of Eq.~\eqref{eq:prob-dist-medium}, one needs to divide by $1-\Delta(0)$.

\begin{figure}
    \includegraphics[width=\columnwidth]{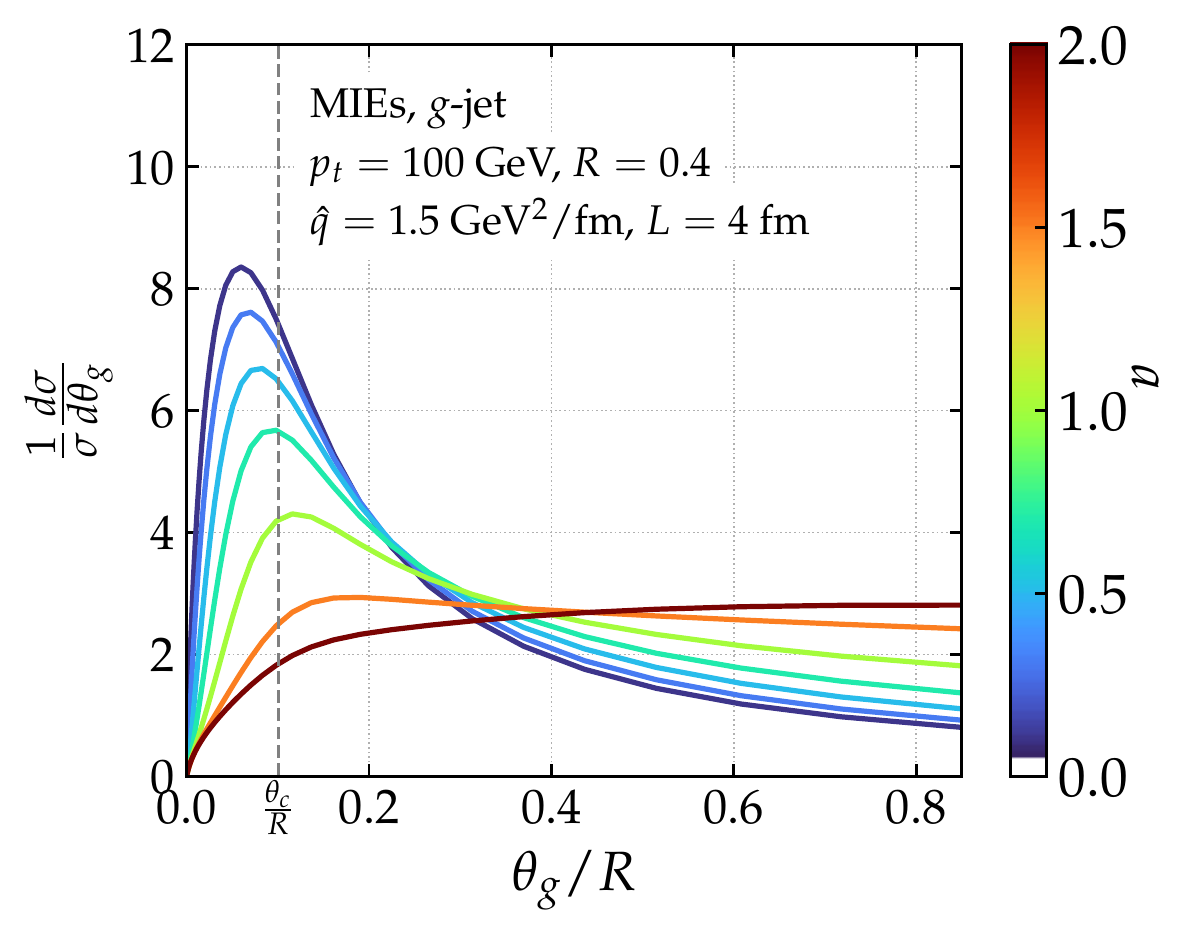}   
    \caption{The $\theta_g$-distribution for gluon jets using only medium induced emissions from Eq.~\eqref{eq:prob-dist-medium}. The small $\theta_g$ behavior is dominated by the broadening term.}
    \label{fig:Pmie}
\end{figure}

In Fig.~\ref{fig:Pmie}, we represent the $\theta_g$-distribution computed with medium induced emissions only. We observe how the small-$\theta_g$ behavior is strongly modified with respect to its vacuum counterpart (see Fig.~\ref{fig:vac-veto-thetag}) in the $a\to 0 $ limit. This behavior is related to the absence of collinear singularities in the medium-induced branching kernel as due to the broadening term. 
 \begin{figure*}
     \includegraphics[width=0.49\linewidth]{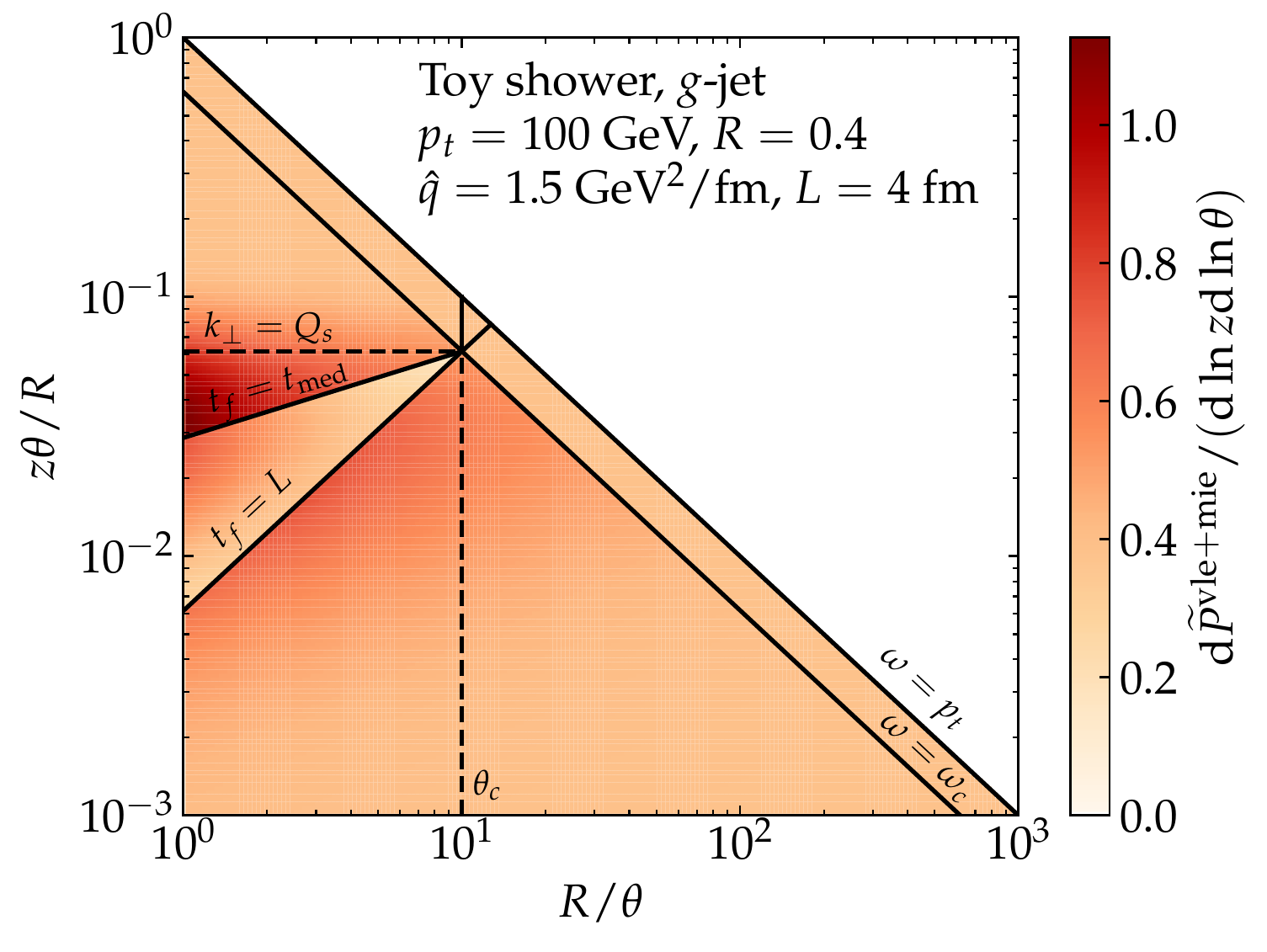}   
     \includegraphics[width=0.49\linewidth]{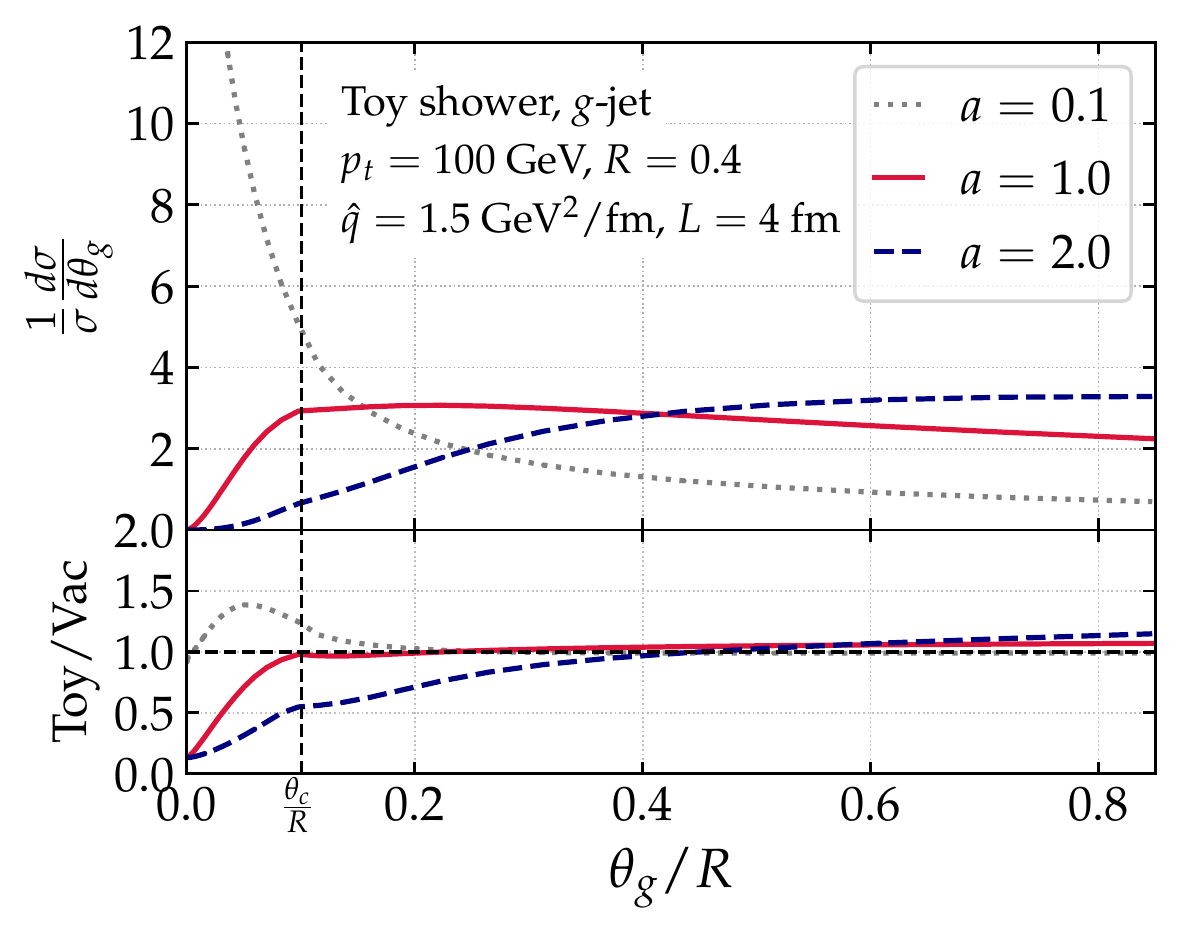}
    \caption{Left: Phase-space density of the branching kernel in the toy shower given by Eq.~\eqref{eq:p-med}. It is interesting to notice that medium induced emissions refill the veto region. Right: The $\theta_g$-distribution for gluon jets using both vacuum-like and medium induced emissions for various values of the DyG parameter $a$. The combined effect of VLE and MIE favors wider emissions.}
    \label{fig:toy-lp-thetag}
\end{figure*}

\paragraph*{Combining vacuum-like and medium-induced emissions.} At this stage, we can construct the probability distribution for a splitting to be the hardest accounting for both vacuum-like and medium induced emissions. Our formula is grounded on the factorization in time between the two types of processes that holds within the DLA. This factorization states that in-medium vacuum-like emissions occur first, in an angular ordered way, followed by time-ordered medium-induced emissions~\cite{Caucal:2018dla}. Further, we impose the following pair of approximations:
\begin{itemize}
\item{Transverse momentum broadening after the emission process is neglected for the in-medium vacuum-like splittings. This would shift the final value of $\theta_g$ by a typical angle of order $Q_s/\omega$ which is indeed negligible in the in-medium region above the line $k_\perp =Q_s$ in Fig.~\ref{fig:medium-lp}. Below this line, this approximation is less justified, and the effect of transverse momentum broadening in jet substructure observables deserves further studies.}
\item{Only relatively hard, primary, medium-induced emissions remains inside the jet cone. In principle, these emissions trigger medium-induced cascades that rapidly develop a turbulent behaviour leading to the multiplication of soft gluons with energy below the multiple branching scale $\omega_{\rm br}\sim\alpha_{s,\rm med}^2\omega_c$~\cite{Blaizot:2013hx,Blaizot:2013vha,Blaizot:2015jea}. Our approximation is valid if these gluons (with $\omega \lesssim \omega_{\rm br}$) are deviated outside the jet cone. Since the typical angle of a gluon in the multiple branching regime is $\theta_{\rm br}\sim (\hat{q}/(\alpha_{s,\rm med}^2\omega^3))^{1/4}$ \cite{Blaizot:2014rla,Blaizot:2014ula}, the condition on the jet radius is $R\lesssim \theta_c/\alpha_{s,\rm med}^2$. For our choice of medium parameters, one gets $ \theta_c/\alpha_{s,\rm med}^2\sim 0.7$ which is indeed larger than the cone sizes studied here.}
\end{itemize}
Under these approximations, the probability distribution for a splitting to be the hardest in the full shower can be written as
\begin{align}
\label{eq:p-med}
\dd^2\mathcal P^{\rm {med}}(z,\theta|a)& =(\dd^2\widetilde{P}^{\rm {vac}}(z,\theta)\Theta_{\notin \textrm{veto}}+\dd^2\widetilde{P}^{\rm {mie}}(z,\theta)) \nn
&\times \Delta_{\notin\rm veto}(\kappa|a)  \Delta^{{\rm mie}}(\kappa|a).
\end{align}
The interpretation of the previous formula is quite transparent from a physical point of view. The tagged splitting can be either a vacuum or a medium induced emission and, for both cases, one has to ensure that emissions of any type with a $\kappa'>\kappa$ are vetoed. We would like to remark that Eq.~\eqref{eq:p-med} can only be taken, at best, as a proxy for a realistic in-medium shower. The Lund plane density of the branching kernels in Eq.~\eqref{eq:p-med}, i.e. $\dd^2\widetilde{P}^{\rm {vac}}(z,\theta)\Theta_{\notin \textrm{veto}}+\dd^2\widetilde{P}^{\rm {mie}}(z,\theta)$, can be found in the left panel of Fig.~\ref{fig:toy-lp-thetag}. In this representation, the vacuum branching kernel is completely uniform except for the fact that it does not populate the veto region. In contrast, due to momentum broadening, medium induced emissions have a typical transverse momentum of $k_\perp\sim Q_s$ in the multiple, soft scattering approximation and thus the enhancement observed in Fig.~\ref{fig:toy-lp-thetag} around this scale.

The difference between the toy shower and the vacuum result is shown in the right panel of Fig.~\ref{fig:toy-lp-thetag}. An interesting point to notice is that the average value of the tagged $\theta_g$ increases when including the medium-induced component. Indeed, at small $\theta_g$, the MIEs cause the distributions to go to zero faster because of the absence of collinear singularity in their emission kernel, leading to a depletion when compared to VLE. The transition angle at which we observe an enhancement depends on the value of $a$: it will be below ($a\ll 1$) or above ($a\sim \mathcal{O}(1)$) the critical resolution angle $\theta_c$. This is reflected on the lower panel of the plot where we clearly observe that the ratio between the toy shower and the vacuum result goes below $1$ at an angle whose value increases with increasing values of $a$.
\subsubsection{Energy loss}
\label{sec:e-loss}
Up to now, we have ignored one of the main distinctive features of in-medium jet propagation, that is, jet energy loss. We have shown in the previous Section that the angular distribution of medium induced emissions is broader than the corresponding vacuum one. Therefore, a `vacuum' jet with a given cone $R$ and transverse momentum $p_{t0}$, will lose energy due to MIEs radiated out of the cone, i.e. with $\theta> R$. The main effect of the large angle energy loss consists in a redistribution of jets with a given value of $z_g$ and $\theta_g$ due to differential energy loss. In other terms, the energy loss by a given jet triggered by a parton with initial transverse momentum $p_{t0}$ depends on the $z_g$ and $\theta_g$ value of the jet after evolution. As the hard spectrum tends to bias towards jet losing less energy than average, this differential energy loss redistributes the amount of jets tagged by $(z_g,\theta_g)$. 

We now write a general formula that encompasses this idea. We call $\mathcal{E}_{i,p_{t0},R}(\varepsilon|z_g,\theta_g)$ the conditional probability for an $i$-initiated jet to radiate energy $\varepsilon$ out of the jet cone $R$, \textit{knowing that} the jet has a dynamically groomed hard branch with kinematic $(z_g,\theta_g)$. Then, the $\theta_g$ distribution for jets having a final transverse momentum $p_t$ is
\begin{align}\label{eq:full-thetag}
 	\left.\frac{1}{\sigma}\frac{\dd \sigma}{\dd\theta_g}\right|_{p_t}&=\mathcal{N}_{\rm med}^{-1}\int \dd\varepsilon\sum_{i\in\{q,g\}}\frac{\dd \sigma_i}{\dd (p_t+\varepsilon)} \nn
	&\times \int \dd z_g\,\mathcal P^{\rm med}_i(z_g,\theta_g)\mathcal{E}_{i,p_t,R}(\varepsilon|z_g,\theta_g)\,,
\end{align}
where $\dd\sigma_i$ is the cross-section for producing a jet with flavor $i$ whose extraction is discussed in App.~\ref{app:spectrum}. Further, $\mathcal{N}_{\rm med}$ is a normalization factor given by
\begin{equation}\label{eq:Nmed}
	\mathcal{N}_{\rm med}(p_t)=\sum_{i\in\{q,g\}}\int \dd\varepsilon\frac{\dd \sigma_i}{\dd (p_t+\varepsilon)}\mathcal{E}_{i,p_{t},R}(\varepsilon)\,,
\end{equation}
since using the law of total probability,
\begin{equation}\label{eq:eloss-total-def}
\mathcal{E}_{i,p_{t},R}(\varepsilon)\equiv\int \dd \theta_g\dd z_g\mathcal{E}_{i,p_{t},R}(\varepsilon|z_g,\theta_g)\mathcal P^{\rm med}_i(z,\theta_g)\,,
\end{equation}
with $\mathcal{E}_{i,p_{t0},R}(\varepsilon)$ the probability for an $i$-jet to lose an energy $\varepsilon$ without any knowledge of its substructure. Notice that we use $\mathcal P^{\rm med}$ in Eq.~\eqref{eq:full-thetag}, i.e. we quench not only vacuum like emissions, but also intrajet medium induced ones. Physically speaking, $\mathcal{N}_{\rm med}$ corresponds to the jet cross section. 

Given the steeply falling nature of the jet spectrum, i.e. $\dd\sigma/\dd p_t\sim p_t^{-n}$ with $n\gg1$, one can write $\dd\sigma/\dd(p_t+\varepsilon)\approx\dd\sigma/\dd p_t\,\exp(-\frac{n\varepsilon}{p_t})$ such that Eq.~\eqref{eq:full-thetag} becomes
\begin{align}\label{full-thg3}
  \left.\frac{1}{\sigma}\frac{\dd \sigma}{\dd\theta_g}\right|_{p_t}&=\mathcal{N}_{\rm med}^{-1}\sum_{i\in\{q,g\}}\frac{\dd \sigma_i}{\dd p_t}\int \dd z_g\mathcal P^{\rm med}_i(z_g,\theta_g)\nn
  &\times\int \dd\varepsilon\,\mathcal{E}_{i,p_{t},R}(\varepsilon|z_g,\theta_g)e^{-\frac{n\varepsilon}{p_t}}\,.
\end{align}
The last line of the previous equation is the Laplace transform of the conditional energy loss probability.

Next, we need to specify the energy loss probability distribution $\mathcal{E}_{i,p_{t},R}(\varepsilon|z_g,\theta_g)$. In the double logarithmic approximation, the jet is dominated by the hardest emission --- the one tagged by dynamical grooming --- and is thus made of two subjets. Neglecting the intrajet multiplicity of the subjets, their energy loss probability can be approximated by that of a single parton\footnote{We do not take into account the fact that the opening angles of the two subjets are different from $R$ and depend on $\theta_g$.} with flavor $i$, out of a cone with opening $R$, denoted by $P^{(1)}_{i,R}(\varepsilon).$. In terms of $P^{(1)}$, the energy loss probability of this two prong system can be written as
\begin{align}
\label{eq:eloss}
&\mathcal{E}_{i,p_{t},R}(\varepsilon|z_g,\theta_g)=\left(1-\Theta_{\rm res}(z_g,\theta_g)\right)P^{(1)}_{i,R}(\varepsilon)+\Theta_{\rm res}(z_g,\theta_g)\nn
&\times\int_0^\infty\dd\varepsilon_1\int_0^\infty\dd\varepsilon_2\,P^{(1)}_{i,R}(\varepsilon_1)P^{(1)}_{g,R}(\varepsilon_2)\delta(\varepsilon-\varepsilon_1-\varepsilon_2)\,,
\end{align}
where the resolution condition reads
\beq
\Theta_{\rm res}(z_g,\theta_g)=\Theta(\theta_g-\theta_c)\Theta(z_g\theta_gp_t-k_{\perp,\rm med})\, ,
\eeq
and it selects splittings in the red region of Fig.~\ref{fig:medium-lp}. Then, Eq.~\eqref{eq:eloss} simply states that if the two prong system is resolved by the medium, the jet energy loss is the sum of the energy losses of each subjet. On the other hand, if the two prongs are not resolved, the full jet loses energy as a single subjet with the color charge of its initiator. Plugging Eq.~\eqref{eq:eloss} into Eq.\,\eqref{full-thg3} yields
\begin{align}
\label{eq:full-eloss}
	&\frac{1}{\sigma}\left.\frac{\dd \sigma}{\dd\theta_g}\right|_{p_t}=\mathcal{N}_{\rm med}^{-1}\sum_{i\in\{q,g\}}\frac{\dd \sigma^h_i}{\dd p_t}\int \dd z_g \mathcal{P}_i^{\rm med}(z_g,\theta_g)\nn
	&\times\left[(1-\Theta_{\rm res})\mathcal Q_i(p_t,R)+\Theta_{\rm res}\mathcal Q_g(p_t,R)\mathcal Q_i(p_t,R)\right]\,,
\end{align}
where we have defined 
\begin{align}
\label{eq:qw-zero}
	\mathcal Q_i(p_t,R)&\equiv\int_0^{\infty}\dd\varepsilon\,P^{(1)}_{i,R}(\varepsilon)\exp\left(-\frac{n\varepsilon}{p_t}\right) \, .
\end{align}
The last step is to find an approximation for the function $P^{(1)}_{i,R}(\varepsilon)$ or equivalently, the quenching weight $\mathcal{Q}_i(p_t,R)$.  Neglecting the intrajet activity of the subjet, $P^{(1)}_{i,R}(\varepsilon)$ can be approximated by the energy loss probability distribution of a single parton of flavor $i$ out of a cone of size $R$. Evaluating the Laplace transform, we arrive to the well known expression for the quenching weight~\cite{Baier:2001yt,Salgado:2003gb}:
\begin{align}
\label{eq:qw}
	&\mathcal Q_i(p_t,R)=\exp\left[\int_R^\infty\rmd\theta\int_0^1\rmd z\frac{\rmd^2 \widetilde{P}^{\rm mie}}{\rmd\theta\rmd z}\left(\rme^{-\frac{n\omega}{p_t}}-1\right)\right]\,.
\end{align}
At this point, an important remark is in order. We have argued that $\dd^2\widetilde{P}^{\rm mie}$ accurately describes the intrajet medium-induced activity. Then, at first glance, it might seem contradictory to use this very same branching kernel to estimate the number of gluons that are deviated outside the jet cone. The physical reason behind this apparent contradiction was presented in Refs.\,\cite{Blaizot:2013hx,Fister:2014zxa} (see Ref.~\cite{Blaizot:2015jea} for a review). It is related to the turbulent behaviour of the medium-induced cascade that efficiently degrades the initial energy into very soft quanta. This turbulent cascade has a fixed point which is identical to the BDMPS-Z spectrum that gives the $z$ dependence of $\dd^2\widetilde{P}^{\rm mie}$ and explains, a posteriori, why Eq.\,\eqref{eq:qw} is a good estimation. 

As argued previously, the typical angle of soft gluons in the multiple branching regime is $\theta_{\rm br}(\omega)$. Therefore, the criterion for a medium-induced gluon to be deviated out of the jet cone is $\theta_{\rm br}(\omega)>R$ or $Q_s/\omega>R$. The latter condition corresponds to the case of a relatively hard emission with $\omega>\omega_{\rm br}$. For the values of $R$ we consider, the second condition overwhelms the first one, so that we can safely approximate the angular dependence of $\rmd^2 \widetilde{P}^{\rm mie}$ by $\delta(\theta-Q_s/\omega)$. Then, the quenching weight reads
\begin{align}\label{eq:Q0}
\ln\mathcal{Q}_i(p_t,R)&=\frac{2\alpha_{s,\rm med}C_i}{\pi}\sqrt{\frac{2\omega_c}{\omega_{\rm max}}}\left(1 -\sqrt{\pi\nu\omega_{\rm max}}\right.\nn
&\times\left.\mathrm{Erf}(\sqrt{\nu\omega_{\rm max}})-e^{-\nu\omega_{\rm max}}\right)\,,
\end{align}
with $\omega_{\rm max}\!=\!\textrm{min}(Q_s/R,\omega_c)$ and $\nu\!=\!n/p_t$. Note that recent works have gone beyond the single parton energy loss picture for global observables by resuming the effects of the fluctuating
substructure on the total energy loss~\cite{Mehtar-Tani:2017web,Mehtar-Tani:2021fud,Takacs:2021bpv}. We will extend that formalism to jet substructure observables in a separate publication~\cite{kt:paper}.

We have checked that this quenching weight gives reasonable values for the $R_{AA}$ ratio of jet cross-sections with our choice of medium parameters. As alluded above, the jet cross-section in Pb-Pb is given by $\mathcal{N}_{\rm med}$, which can be obtained either from Eqs.\,\eqref{eq:Nmed}-\eqref{eq:eloss-total-def} or from Eq.\,\eqref{eq:full-eloss} thanks to the self-normalization of the $\theta_g$ distribution. In both cases, we observe a mild $a$-dependence of this jet cross-section, as a consequence of the main underlying approximation of Eq.\,\eqref{eq:eloss}, namely the fact that we neglect the vacuum-like multiplicity of the resolved or unresolved subjets. This uncertainty in the jet cross-section is harmless for the shape of the $\theta_g$ distributions, which are self-normalized by construction.

\begin{figure}
     \includegraphics[width=\columnwidth]{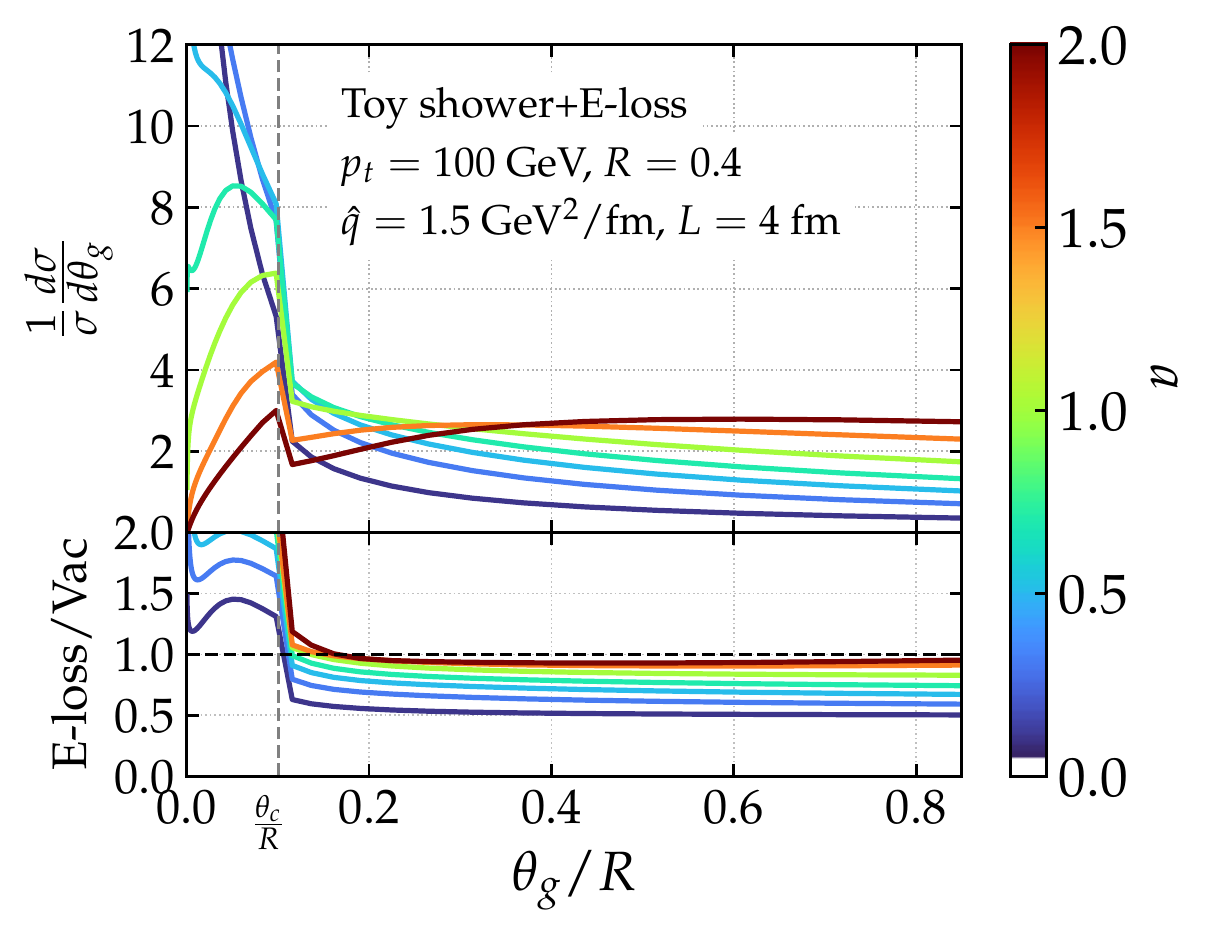}
    \caption{The $\theta_g$-distribution obtained with Eq.~\eqref{eq:full-eloss}. Including differential energy loss results into a sharp transition in the distributions at $\theta_c$.}
    \label{fig:eloss}
\end{figure}

In Fig.~\ref{fig:eloss} the quenched $\theta_g$-distributions are displayed. The most remarkable feature of these distributions is the keen transition at $\theta_g\!=\!\theta_c$. This arises due to several reasons that we proceed to analyse. To start with, our energy loss model, i.e. Eq.~\eqref{eq:eloss}, contains a sharp distinction between resolved and unresolved splittings that translates into branchings with $\theta \geq \theta_c$ loosing more energy than those with $\theta \leq \theta_c$. Then, the steeply falling nature of the spectrum drastically reduces the possibilities of these wide angle branchings that lost a substantial amount of energy to end up in the selected $p_t$-window. That is, the least quenched jets, i.e. those splittings with $\theta \leq \theta_c$ (the green region in Fig.~\ref{fig:medium-lp}), dominate the $\theta_g$-distribution and therefore a narrowing is expected. This is a well known effect typically referred to as `selection bias' or `filtering effect', see e.g. Ref.~\cite{Brewer:2018dfs,Du:2020pmp,Du:2021pqa,Takacs:2021bpv} for a possible way out. This feature also explains the $a$-dependence of the ratio in the bottom panel. In fact, we have shown in Eq.~\eqref{eq:theta-cut} that the maximum of the $\theta_g$-distribution for vacuum splittings is inversely proportional to $a$. This estimate is still correct when including medium induced emissions. Therefore, when $a\geq 1$ the probability of tagging a wide splittings is larger than selecting a narrow one. However, once energy loss is included, those very few narrow splittings will lose significantly less energy than their wide angle counterparts. Consequently, their probability is enhanced with respect to vacuum and the ratio goes above one. On the other hand, when $a\leq 1$ small angle splittings are typically selected. Following the same reasoning, the very few large angle splittings will be even more suppressed than in vacuum due to incoherent energy loss. 

Finally, we would like to comment on the effect of the quark/gluon fraction resulting from the sum over flavors in Eq.\,\eqref{eq:full-eloss}. In the vacuum, we expect quark initiated jets to have a narrower $\theta_g$ distribution than gluon initiated jets (see e.g.\ Eq.\,\eqref{eq:theta-cut}). Therefore, the $\theta_g$ distribution is also sensitive to the different quark-gluon fraction of the hard spectrum in Pb-Pb collisions compared to $pp$, but as shown in Appendix~\ref{app:spectrum}, the overall effect is very mild. That said, once large angle jet energy loss is included, since quark jets lose less energy than gluon jets, we expect a filtering effect towards quark initiated jets, leading to an even narrower $\theta_g$ distribution. This effect is accounted for in our analytic calculation and in Fig.~\ref{fig:eloss}. To disentangle these two filtering effects, i.e. (i) towards coherent, "unresolved" jets and (ii) quark-initiated jets, an interesting possibility is to measure the $\theta_g$ distribution in Z/$\gamma$+jet events~\cite{Brewer:2018dfs,Takacs:2021bpv,Brewer:2021hmh}.

\subsubsection{Path-length fluctuations}
\begin{figure*}
     \includegraphics[scale=0.7]{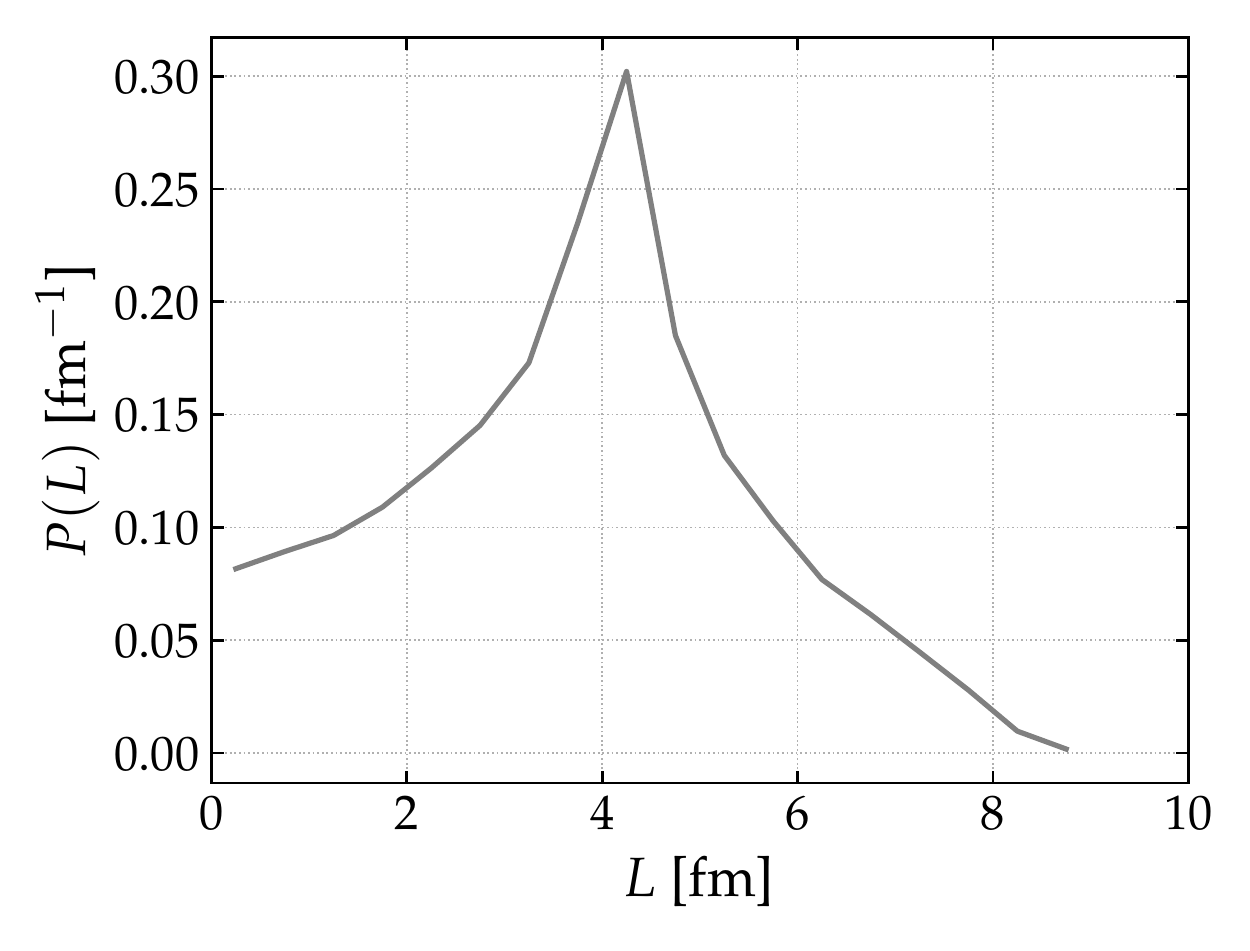} \includegraphics[scale=0.7]{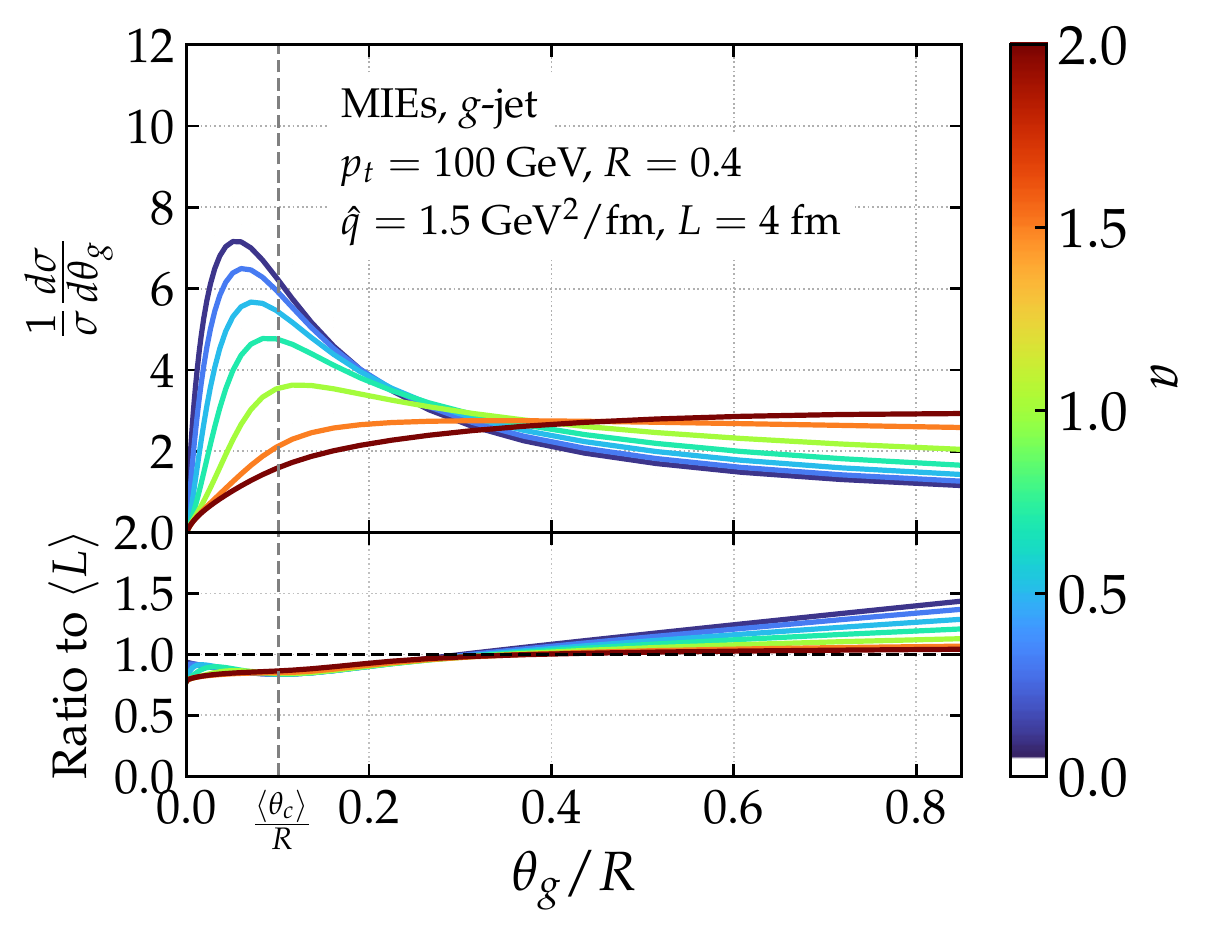}
    \caption{Left: Probability distribution of the fluctuating jet path length in a simplified scenario described in the main text. Right: The $\theta_g$-distribution for various values of the Dynamical Grooming parameter $a$ when including medium induced emissions only and path-length fluctuations. The fluctuations smoothen the sharp transition around $\theta_c$. The asymmetric nature of the jet path length distribution results into a non-flat ratio.}
    \label{fig:lfluc}
\end{figure*}
So far, we assumed that the medium is a homogeneous brick of fixed length. This ignores the rapid expansion of the medium and oversimplifies the geometry of a heavy-ion collision where the hard scattering that produces the jet can take place anywhere inside the geometric overlap area between the two colliding nuclei. We expect these fluctuations to smoothen our sharply peaked curves around the critical angle $\theta_c$. In this section we introduce an extension of the brick model to capture some of these fluctuations.  

Previous studies showed that a simple Bjorken-like expansion of the medium is well captured by rescaling the jet quenching parameter $\hat q$ of a homogeneous brick \cite{Adhya:2019qse,Caucal:2020uic}, $\hat{q}\equiv ||\hat{q}(t)||_{1/2}$, where $|| f(t)||_{1/2}$ stands for the $1/2$-norm of the function $f(t)$ with compact support. This scaling is a consequence of the local nature of the medium-induced emissions in the multiple soft scattering regime $\omega\ll\omega_c$ \footnote{This scaling is therefore distinct from the one discovered in Ref.~\cite{Salgado:2002cd} that works for processes dominated by the most energetic medium-induced emissions ($\omega\sim\omega_c$). As shown in Ref.~\cite{Caucal:2020uic}, it is also violated by VLEs via a change of the phase space boundaries that we neglect in this study.}. Since we do not consider medium-induced emissions harder than $\omega_c$, we invoke this scaling to extend our results to the Bjorken expansion case.

To capture the fluctuation in the path length of the jet for central collisions, we propose the following model: (i) the interaction region is approximated by a circle of radius $R=4$ fm around the center of the collision, (ii) random $(x,y)$ coordinates of hard scatterings are sampled uniformly in the interaction region, (iii) each creation point is connected with a hard-scattering leading-order matrix element from {\tt Pythia8} \cite{Sjostrand:2014zea} (Monash13 tune \cite{Skands:2014pea}) in proton-proton collisions, assigning the 4-momenta of the outgoing legs and (iv) the path lengths are determined by the intersection of the path with the edge of the interaction region. The distribution of the resulted path lengths is shown in the left panel of Fig.~\ref{fig:lfluc}, centered around 4 fm, however, $\langle L\rangle=3.75$ fm due to the asymmetry of the distribution. Even though this model is over-simplistic, it is sufficient to qualitatively understand the effects of the path length fluctuations on the $\theta_g$ distribution. More precise phenomenology would require to account for the nuclear thickness function and the precise shape of the interaction region across various centrality classes.

The $\theta_g$-distribution obtained with the medium-induced branching kernel, see Eq.~\eqref{eq:prob-dist-medium}, and a fluctuating path length is presented in the right panel of Fig.~\ref{fig:lfluc}. The ratio to the average $L$ result is displayed in the bottom panel. The enhancement of large $\theta_g$ values is rooted in the asymmetric nature of the path-length distribution, see left panel of Fig.~\ref{fig:lfluc}. More concretely, shorter than average path-lengths are more probable. This automatically translates into a distribution of $\theta_c$ values that tend to be larger than average due to the $\theta_c\propto L^{-3/2}$ scaling. Consequently, the $\theta_g$ distribution gets broader when path length fluctuations are included. 
\subsection{Final theoretical results}
\label{sec:final-theory}
\begin{figure*}[hbt]
     \includegraphics[width=\textwidth]{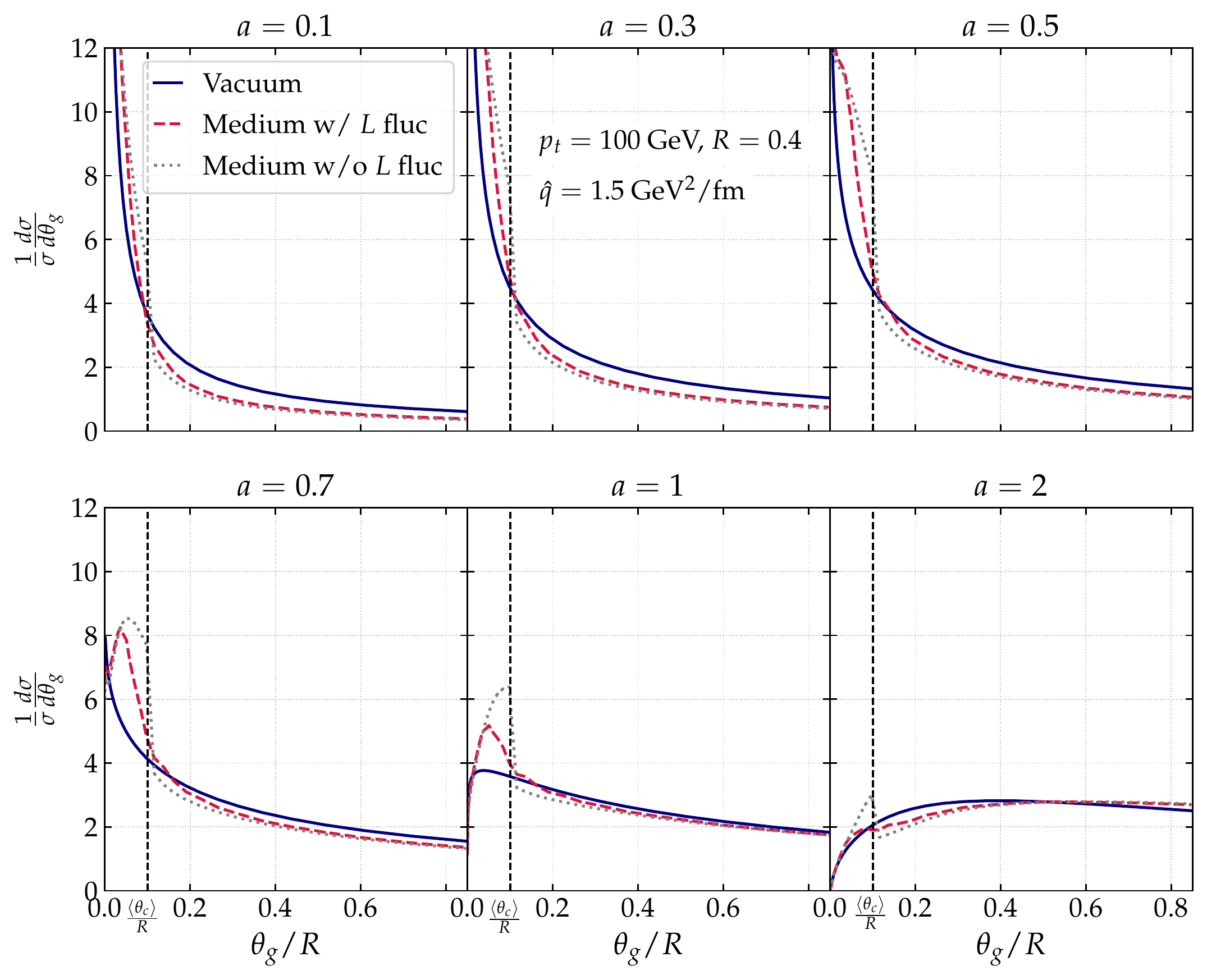}
    \caption{The $\theta_g$-distribution for the toy model given by Eq.~\eqref{eq:full-eloss} for various values of the DyG parameter $a$ in vacuum (solid, blue) and in the medium with (dashed, red) and without (dotted, gray) path length fluctuations. The jet  path length fluctuations do not wash out the peak around $\theta_c$. }
    \label{fig:final-theory}
\end{figure*}

Finally, we present our theoretical curves including all the ingredients discussed in the previous paragraphs in Fig.~\ref{fig:final-theory}. Compared to Fig.~\ref{fig:eloss}, we observe that the main effect of introducing the path length fluctuations is to smoothen the transition around the critical angle $\theta_c$. Consequently, the peak of the medium modified $\theta_g$ distribution is shifted towards slightly smaller values of the opening angle. It easy to observe by eye that there are choices of the Dynamical Grooming parameter $a$ which enhance the relative difference between the medium and the vacuum $\theta_g$ distributions. For $a\sim 2$, we do not see a significant deviation, whereas values of $a$ close to 1 give a pronounced peak around the mean value of $\theta_c$ which is not present in the vacuum distribution. Therefore, we expect that measuring the dynamically groomed jet radius with $a\sim 1$ will provide a clear evidence of the existence of a characteristic (de)coherence angle.

\begin{figure}
     \includegraphics[width=\columnwidth]{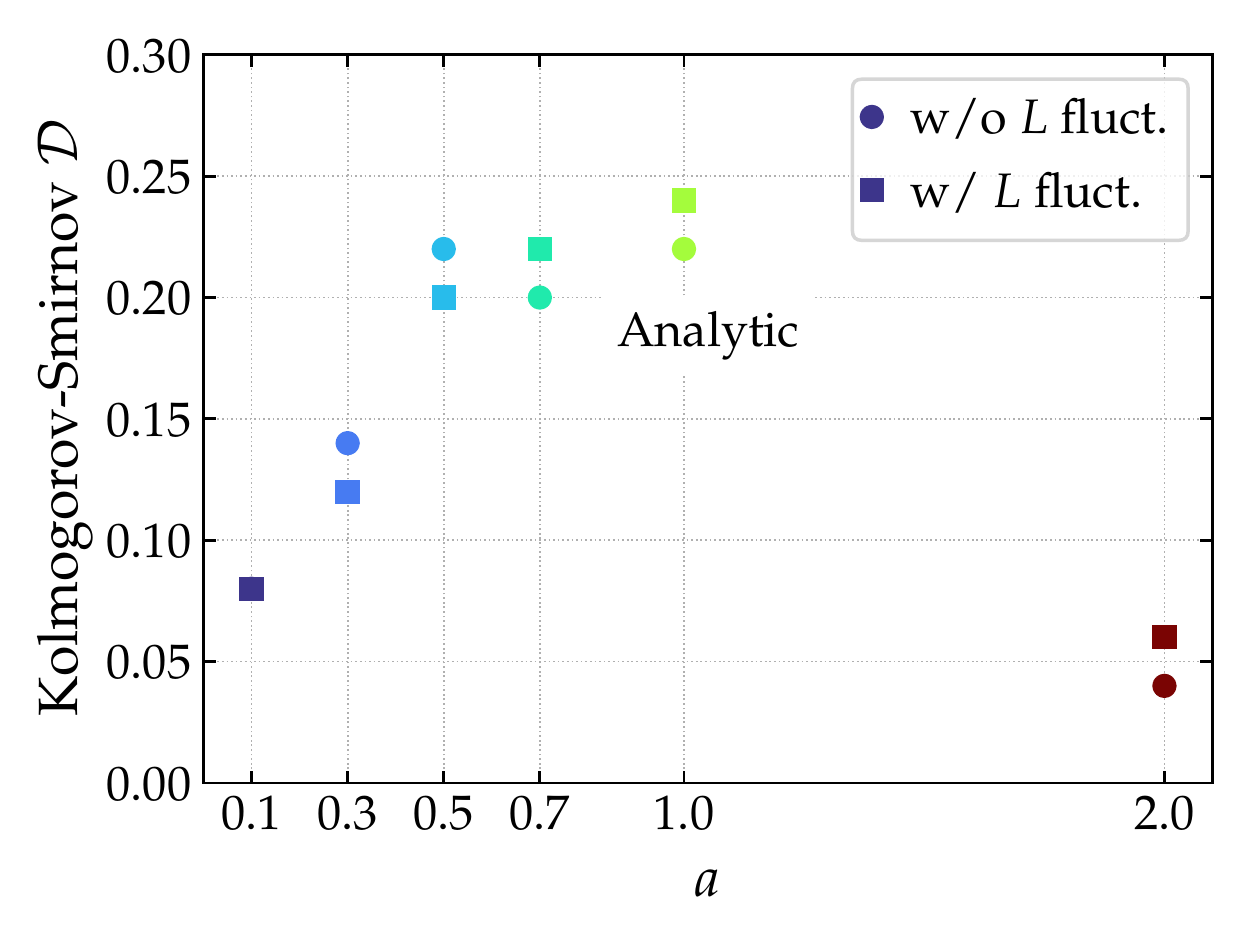}
    \caption{The Kolmogorovs-Smirnov distance defined in Eq.~\eqref{eq:ks-distance} as a function of the Dynamical Grooming parameter $a$ for the theoretical results with (squares) and without (circles) jet path length fluctuations. The bigger the KS value, the easier to discriminate between vacuum and medium distributions.}
    \label{fig:ks-test-theory}
\end{figure}

In order to gauge the sensitivity of the $\theta_g$ observable to medium physics in a more quantitative way, we choose the Kolmogorov-Smirnov metric as a measure of the differences between the vacuum and in-medium distributions. The Kolmogorov-Smirnov (KS) distance\footnote{The authors would like to thank Marta Verweij for suggesting this metric in a different context.}, $\mathcal{D}$, is defined as
\begin{equation}
\label{eq:ks-distance}
\mathcal{D}=\underset{0\le\theta_g\le R}{\rm{max}}|\Sigma^{\rm vac}(\theta_g)-\Sigma^{\rm med}(\theta_g)|,
\end{equation}
where $\Sigma$ denotes the cumulative distribution
\begin{equation}
\Sigma(\theta_g)=\displaystyle\int_0^{\theta_g}\dd\theta' \displaystyle\frac{1}{\sigma}\displaystyle\frac{\dd \sigma}{\dd\theta'}\, .
\end{equation}
That is, the KS metric corresponds to the maximal distance between the cumulated spectra. The largest $\mathcal{D}$ is, the most distinct the two distributions are and, consequently, the larger the discriminating power of $\theta_g$ is. We choose to use this more involved metric instead of the usual ratio because of the strong differences in shape between the medium and vacuum distributions. In Fig.\,\ref{fig:ks-test-theory}, we display the value of the Kolmogorov-Smirnov distance resulting from our analytic calculation of the $\theta_g$ distribution for several values of the grooming parameter $a$. The largest distance corresponds to $a=1$, but we also observe that smaller values, between $0.5$ and $1$ gives also a large $\mathcal{D}$ value. The other interesting feature of this plot, that we shall also observe in Monte-Carlo simulations, is the reduction of the Kolmogorov-Smirnov distance once path length fluctuations are included, as a consequence of smoother transition between coherent and incoherent subjet energy loss.

To summarise and conclude this analytic section, we emphasise that our pQCD motivated theoretical model, which relies on the factorisation between VLEs and MIEs in DLA and the multiple soft scattering approximation, predicts a significant modification of the $\theta_g$ distribution around the critical angle $\theta_c$, as a consequence of multiple imprints of different physical mechanisms, whose dominant one is the filtering effect towards fully coherent jets. The second important result is that, by pursuing an analytic approach, we are able to provide theoretical guidance on the optimal values of the DyG parameter $a$ that maximise the discrimination power of this observable in order to probe the physics of color (de)coherence experimentally. The alluded $a$-range is $0.5\lesssim a\lesssim 1$.
 
\section{Monte-Carlo simulations}
\begin{figure*}[hbt]
     \includegraphics[width=\textwidth]{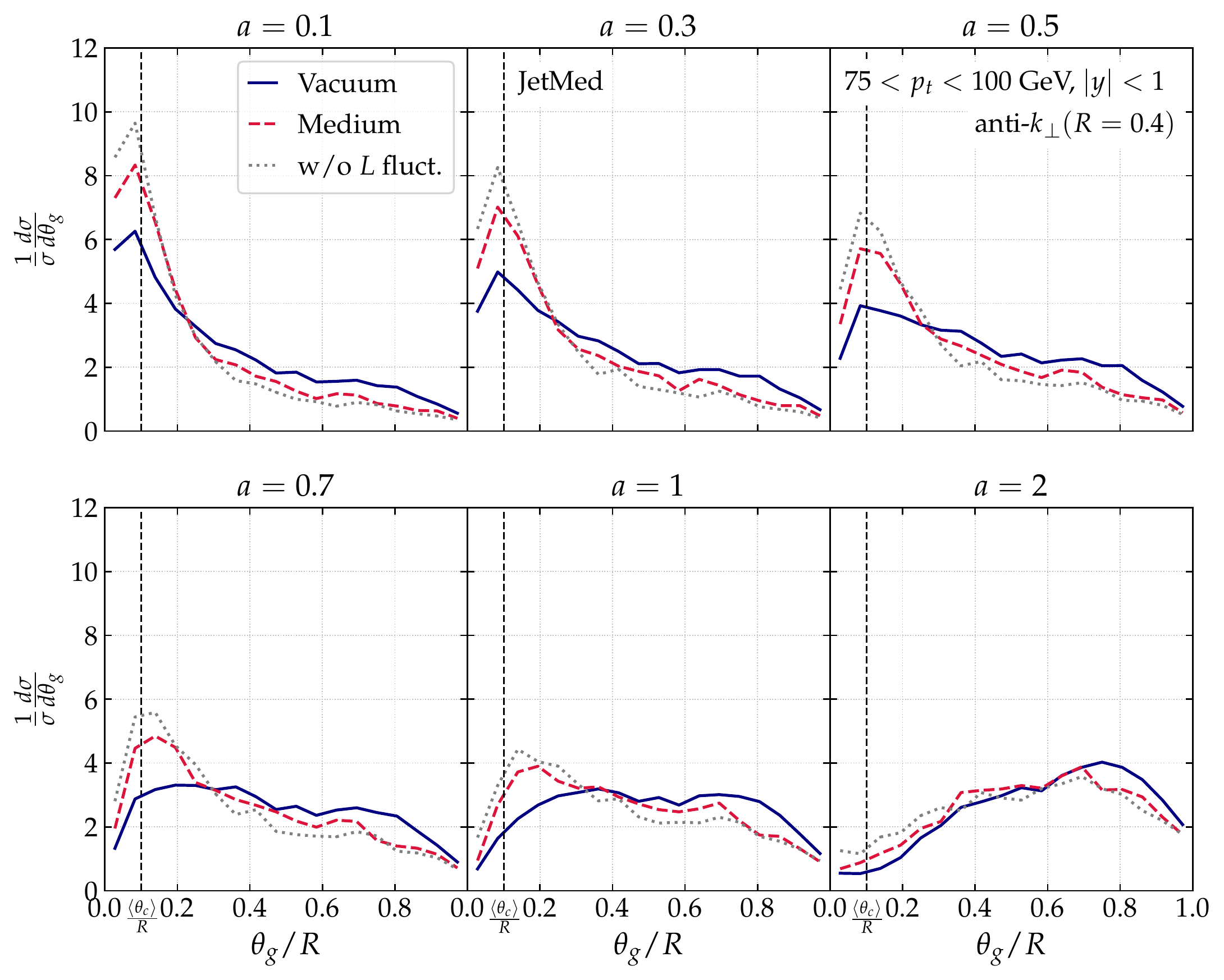}
    \caption{The $\theta_g$-distribution for various values of the DyG parameter $a$ with {\tt{JetMed}} in vacuum (solid, blue) and in the medium with (dashed, red) and without (dotted, gray) jet path length fluctuations. The medium distribution sticks out the flat vacuum benchmark.}
    \label{fig:jetmed}
\end{figure*}
In this section, we numerically explore the $\theta_g$ distribution with three state-of-the-art jet quenching Monte-Carlo codes: JetMed~\cite{Caucal:2018dla}, the Hybrid model~\cite{Casalderrey-Solana:2014bpa} and Jewel~\cite{Zapp:2012ak,Zapp:2013vla}. In all cases, we generate dijet events at $\sqrt s = 5.02$~TeV in Pb+Pb collisions~\footnote{Actually, we only generated the JetMed events by ourselves. We have obtained the Jewel samples from \url{https://jetquenchingtools.github.io}, while the Hybrid events have been kindly provided by Daniel Pablos.}. For each event, particles are clustered on an event-by-event basis into anti-$k_t$ jets~\cite{Cacciari:2008gp} with $R=0.4$ and re-clustered with the Cambridge/Aachen~\cite{Dokshitzer:1997in} algorithm to obtain an angular ordered clustering sequence. The analysis is performed on jets with transverse momenta $75\!<\!p_t\!<\!100$~GeV and rapidities $|y|\!<\!1$. 
\begin{figure*}
     \includegraphics[scale=0.7]{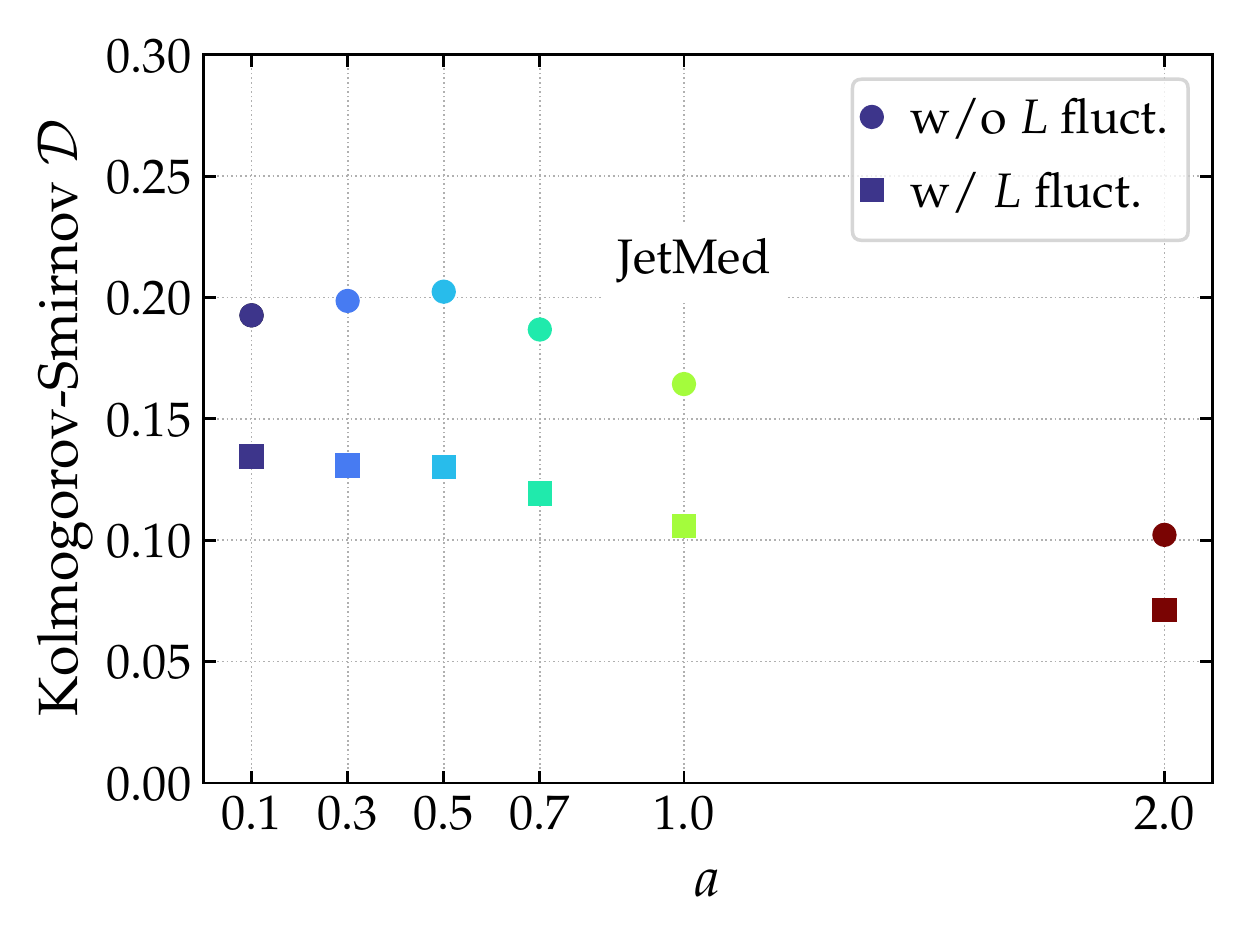}\includegraphics[scale=0.7]{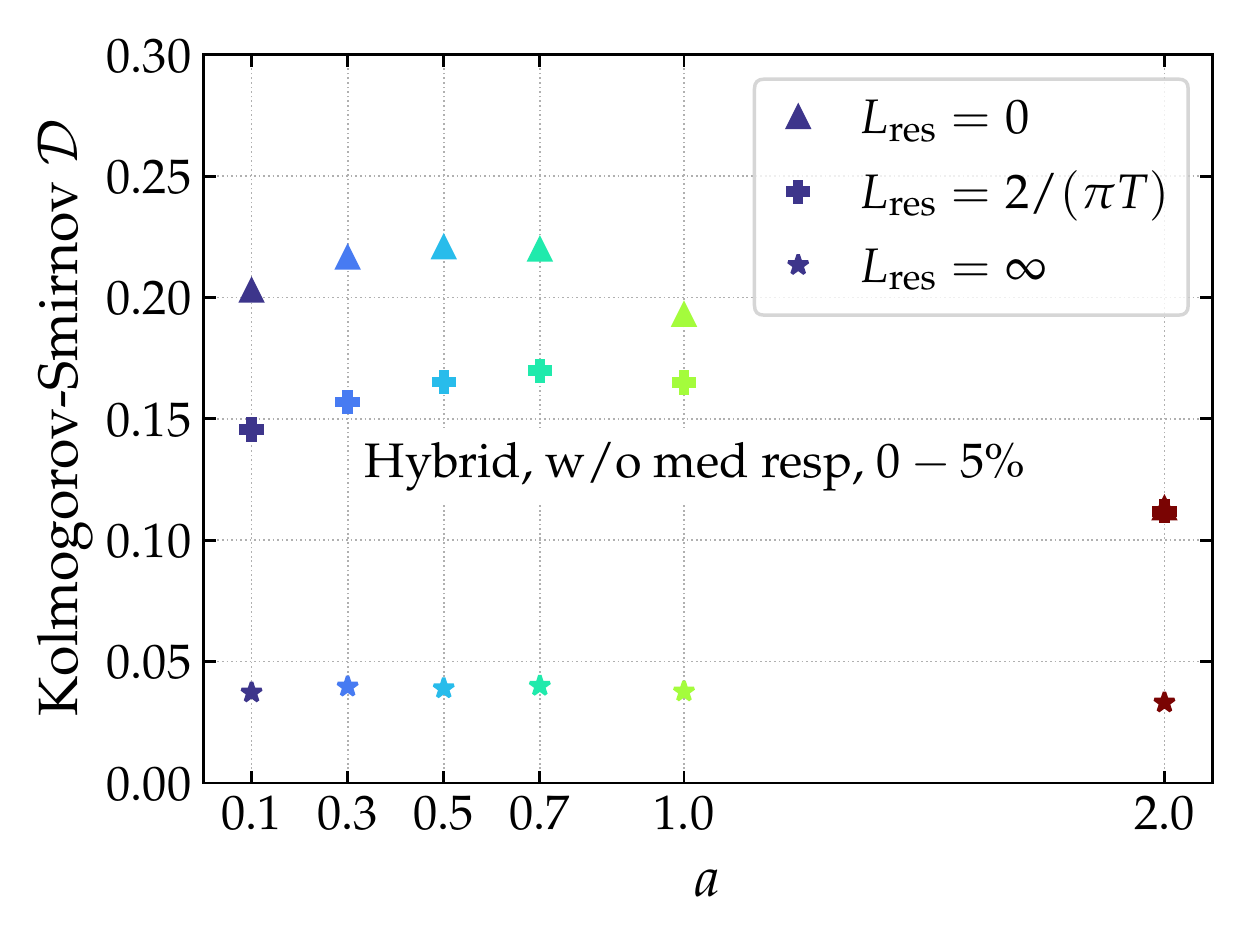}
    \caption{Kolmogorov-Smirnov distance defined in Eq.~\eqref{eq:ks-distance} as a function of the Dynamical Grooming parameter $a$ with the JetMed parton shower (left) and the {\tt Hybrid} model (right). Weak and strong couplingg description of the jet-medium interaction are in qualitative agreement.}
    \label{fig:jetmed-hybrid-ks-test}
\end{figure*}
\subsection{{\tt{JetMed}}}
\label{sec:jetmed}
To begin with, we present results for the Monte-Carlo framework that is closest in spirit to the semi-analytic model presented in the previous section. The Monte-Carlo {\tt{JetMed}} is a parton shower based on the factorization between vacuum-like emissions and medium-induced emissions that holds in the double-logarithmic approximation for the former and multiple soft scattering approximation for the latter. The main differences to be expected between the analytic approach and the numerical results concern (i) the inclusion of part of the single logarithmic corrections to the vacuum-like shower, through the running of the QCD couplingg and the hard collinear emissions, (ii) the proper resummation of the medium-induced emissions with formation time $t_{f}^{\rm med}\ll L$ in the multiple branching regime, (iii) the quenching weight approximation is relaxed since the jet energy loss is provided for free in a parton shower approach and (iv) transverse momentum broadening after emission that leads to a shift in the final $\theta_g$ value of the tagged subjet is accounted for. Notice that we have extended the original code to include jet path length fluctuations using the same model of the geometry as in the analytics. 

The resulting $\theta_g$-distributions for {\tt{JetMed}} are displayed in Fig.~\ref{fig:jetmed}. Let us start the discussion with the vacuum curves. A clear quantitative discrepancy exists between them and the analytic ones presented in Fig.~\ref{fig:vacuum-thetag}. We have tested that the main source of this difference is the choice of fixed couplingg in the analytic result. Despite being an easy element to introduce in the analytic calculation, we opt not do it following the logic of this paper that presents results at double logarithmic accuracy. Regarding the medium, we observe the same trends as in the analytic calculation: for $a\leq 1$ they are strongly peaked at the average critical angle $\theta_c$, the relevant angular scale in the problem. The sharpness of the transition is reduced by the inclusion of jet path length fluctuations. An important observation is that the medium curves are not merely shifted towards smaller angles, but actually the whole shape of the distribution is different from the vacuum one. This is quantified by the Kolmogorov-Smirnov distance that we show in the left panel of Fig.~\ref{fig:jetmed-hybrid-ks-test}. Through this metric we confirm that in order to enhance the sensitivity to jet quenching effects a rather small value of $a$ must be chosen. We would like to remark that this optimisation exercise is relatively simple in the case of Dynamical Grooming given that it has a single free parameter and, as noted in Ref.~\cite{Mehtar-Tani:2019rrk}, the $\theta_g$-distribution is invariant under the $a\to1/a$ transformation. This last point immediately reduces the range of $a$-values to scan. Obviously, one could also calculate $\mathcal{D}$ in the two-dimensional parameter space spanned by the SoftDrop condition. However, not only the increased dimensionality but also the possible degeneracy between pairs of ($z_{\rm{cut}},\beta$) complicate the analysis.  

\subsection{Strong vs Weak couplingg approach}
\label{sec:othermcs}
\begin{figure*}
\includegraphics[width=\textwidth]{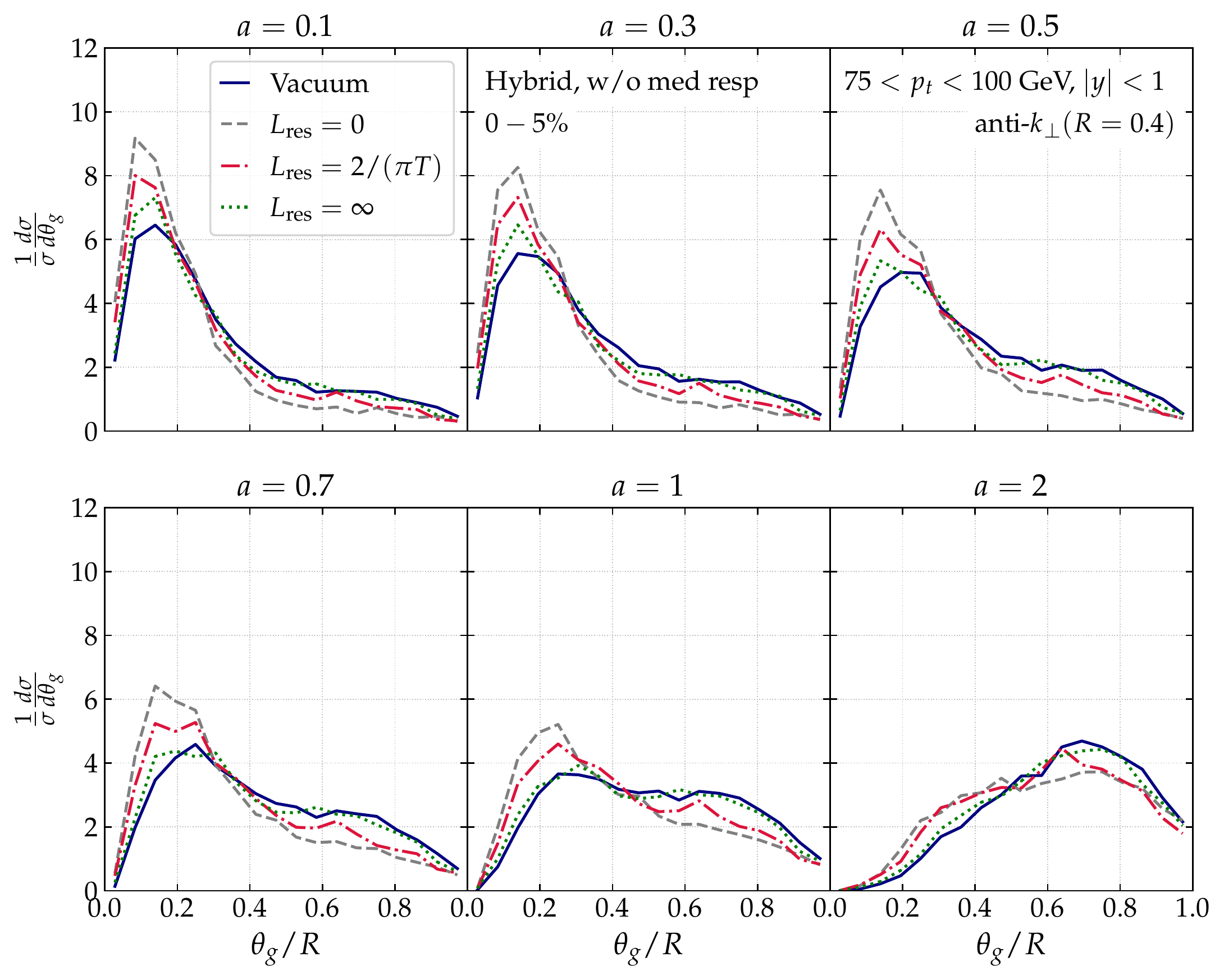}
\caption{The $\theta_g$-distribution for various values of the DyG parameter $a$ with the {\tt{Hybrid}} and no medium response. The fully coherent energy loss ($L_{\rm res}=\infty$) scenario closely resembles the vacuum distribution.}
\label{fig:hybrid-nr}
\end{figure*}
All along this manuscript, we have considered a very specific model of the in-medium shower rooted on two basic pillars: a weak couplingg description of the medium and the multiple soft scattering approximation. The natural question now is whether the observed features of the $\theta_g$-distribution arise only on this concrete picture or are general enough such that any model with some notion of an angular dependence in the energy loss formulation leaves the same footprints in the $\theta_g$-distribution. To address this point we make use of the {\tt{Hybrid}} model. It is beyond the scope of this paper to provide a thorough description of all the ingredients assembled in this code, but we would like to highlight some of the differences in the phase-space of emissions with respect to the discussion surrounding Fig.~\ref{fig:medium-lp}. 

In the strong coupling description of the medium, the existence of a finite resolution length was first considered in Ref.~\cite{Hulcher:2017cpt}. However, this property does not appear naturally as it is the case in the weak-couplingg scenario, but rather has to be introduced by hand as a free parameter called $L_{\rm {res}}$ in the {\tt{Hybrid}} model. From a physics point of view, $L_{\rm {res}}$ is inversely proportional to the local temperature of the medium, the only scale in the problem, and the coefficient of proportionality cannot be computed from first principles but only estimated. In this work we explore three different values of $L_{\rm {res}}$. The fully incoherent case is considered by setting $L_{\rm {res}}=0$. The opposite scenario in which the jet is treated as a single object corresponds to $L_{\rm{res}}=\infty$. Finally, we take an intermediate value of $L_{\rm {res}}=2/(\pi T)$, with $T$ being the local temperature of the plasma. Another important difference with respect to {\tt{JetMed}} is the fact that no-medium induced branching kernel exists in the {\tt{Hybrid}} model, i.e. the splitting probability is the same as in vacuum. On top of that, we consider hadronized samples and switch off the medium response for the purpose of this section. 

We show the dynamically groomed jet radius distributions in Fig.~\ref{fig:hybrid-nr} for the aforementioned values of the screening length $L_{\rm {res}}$. There are two cases for which no angular scale is present in the energy loss mechanism: $L_{\rm{res}}=0$ and $L_{\rm{res}}=\infty$. Clearly, the $L_{\rm{res}}=\infty$ results closely resemble the vacuum distributions. The interpretation of this result is quite transparent: in the $L_{\rm{res}}=\infty$ case only the parent parton loses energy and therefore the filtering effect is drastically reduced. Naturally, the orthogonal scenario where all individual splittings are resolved by the medium, i.e.  $L_{\rm{res}}=0$, exhibits the biggest modification with respect to the vacuum benchmark. The intermediate case of $L_{\rm {res}}=2/(\pi T)$ is in quantitative agreement with the {\tt{JetMed}} results. This allows us to conclude that the proposed observable is agnostic to the fine details of the energy loss mechanism or the physical origin of the resolution scale. If measured experimentally, this observable has enough discriminatory power to discard (or confirm) a fully coherent scenario ($L_{\rm{res}}=\infty$). However, it would be hard to disentangle between the $L_{\rm {res}}=0$ and $0\leq L_{\rm {res}}\leq\infty$ cases. To that end, further constraints on the model from other observables are required, e.g. fit the parameters to describe the $R_{AA}$ and predict the $\theta_g$-distribution. 

\begin{figure*}
    \includegraphics[scale=0.7]{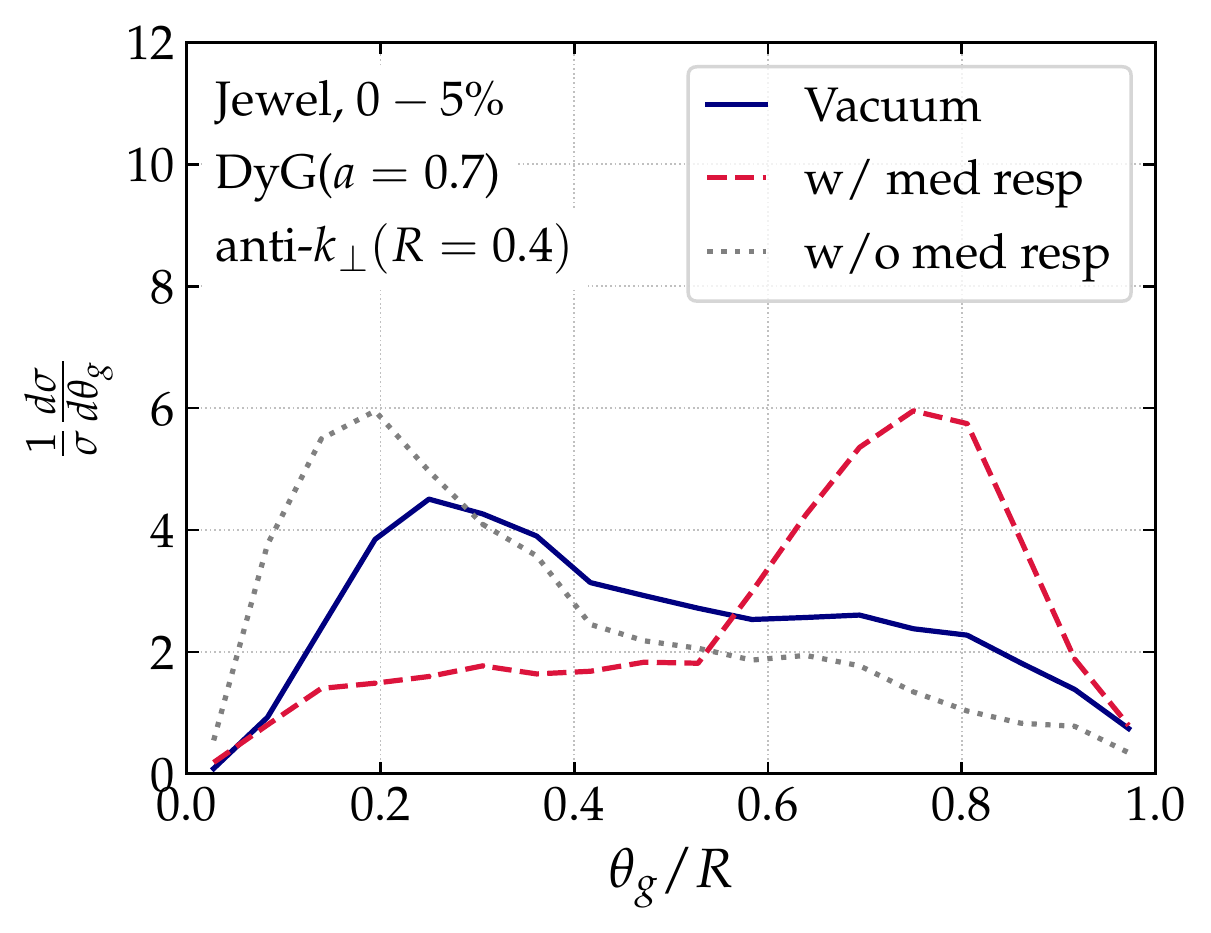}\includegraphics[scale=0.7]{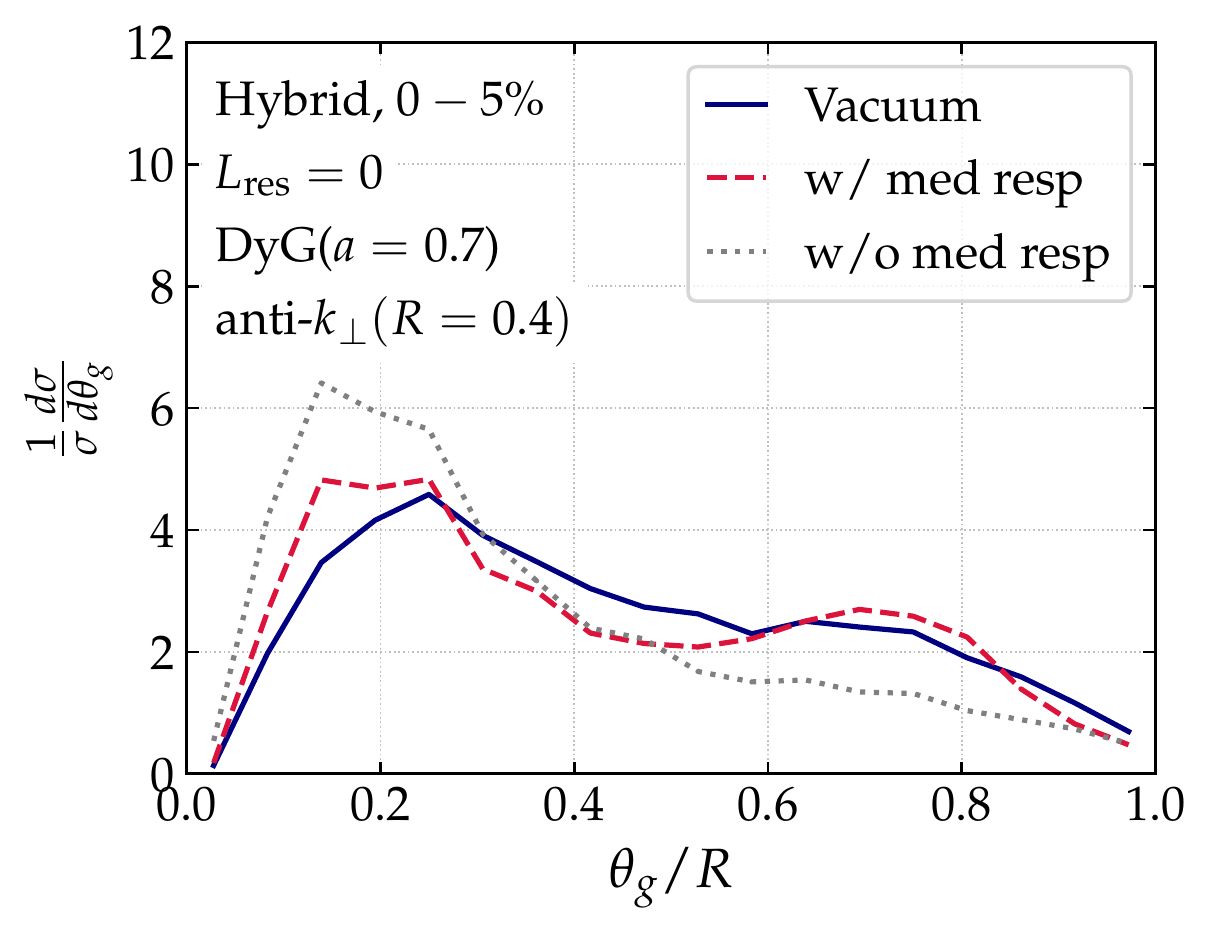}
    \caption{The $\theta_g$-distribution for the {\tt{Jewel}} (left) and the {\tt{Hybrid}} (with $L_{\rm{res}}=0$) models including medium response for $a=0.7$ in the DyG condition. The medium response introduces an enhancement of partons towards wider angles.}
    \label{fig:med-response}
\end{figure*}

Notice, that, in contrast to {\tt{JetMed}}, this Monte-Carlo uses a realistic hydrodynamical profile. Remarkably, the narrowing of the distribution is not washed out by the medium expansion and its geometrical fluctuations. Regarding the impact of hadronisation, previous papers on DyG observables showed that it affects mainly the setups in which $a\leq 1$. This is confirmed by the values of the Kolmogorov-Smirnov metric shown in the right panel of Fig.~\ref{fig:jetmed-hybrid-ks-test}. Indeed, we observe that the optimal value of $a$ shifts from $a\leq 0.5$ in {\tt{JetMed}} to $0.5\leq a \leq 1 $ in the {\tt{Hybrid}} model.

\subsection{Impact of medium response}
\label{sec:medium-response}
The final effect that we would like to quantify is the impact of the medium back reaction in our proposed observable. To that end, we make use two independent models of medium response: the ones implemented in the {\tt{Jewel}} and {\tt{Hybrid}} models. Technically, in order to avoid double counting of the thermal particles momenta, we implemented the `\textit{GridSub1}' method described in Ref.~\cite{KunnawalkamElayavalli:2017hxo} for {\tt{Jewel}} and the background subtraction technique presented in the Appendix A of Ref.~\cite{Casalderrey-Solana:2019ubu} in the {\tt{Hybrid}} case. Further, in both models hadronisation is switched on. One last remark before presenting the results is that in {\tt{Jewel}} the radiation for unresolved emissions is not considered. Then, there is no coherence angle $\theta_c$ in this model and the $\theta_g$-distribution would be mainly sensitive to the filtering effect due to the finite size of the medium. In some sense, it's closest to the $L_{\rm{res}}=0$ in the {\tt{Hybrid}} calculation that we have presented above. 

In Fig.~\ref{fig:med-response} we present the $\theta_g$-distributions for the optimal value of $a$ according to the Kolmogorov-Smirnov tests performed in the previous section. Further, in the case of the {\tt{Hybrid}} model, we fix $L_{\rm{res}}=0$ given that it showed the biggest difference with respect to the vacuum baseline. First of all, the vacuum curves between the two models are in quantitative agreement as expected since they are both based on the Pythia parton shower. The quenched curves without medium response of the two jet quenching Monte-Carlo generators also match. However, once medium response is taken into account the two results differ. Qualitatively, an enhancement of wide-angle splittings is observed. Since particles originated from the medium-back reaction are inherently soft, they can only affect this DyG observable if they appear at large enough angles. Their contribution is sizeably different in the explored models. More concretely, in the {\tt{Jewel}} case medium response completely distorts the shape of the distribution and creates a bump at the edge of the jet cone, while in the {\tt{Hybrid}} case the impact is more moderate but brings the medium and vacuum distributions closer. The signal created by the wake particles clearly pollutes the interpretation of the KS distance in terms of probing $L_{\rm{res}}$. Turning the argument around, these results suggest the potential of this observable to discriminate between different models of the medium back-reaction. However, the large-angle domain is also contaminated by the fluctuating underlying event in heavy-ion collisions, as was shown in Ref.~\cite{Mulligan:2020tim}. In this work, we are interested in designing a pQCD-dominated observable and therefore we proceed to present two possible ways of reducing the impact of soft physics, i.e. both the medium back-reaction and the thermal background: (i) using smaller jet radii and (ii) selecting semi-peripheral events~\footnote{Another alternative that would effectively reduce the impact of medium response, but we do not pursue in this work, would be to increase the jet $p_t$ in the selection.}.  
\subsubsection{Jet radius dependence}
\begin{figure*}
    \includegraphics[scale=0.7]{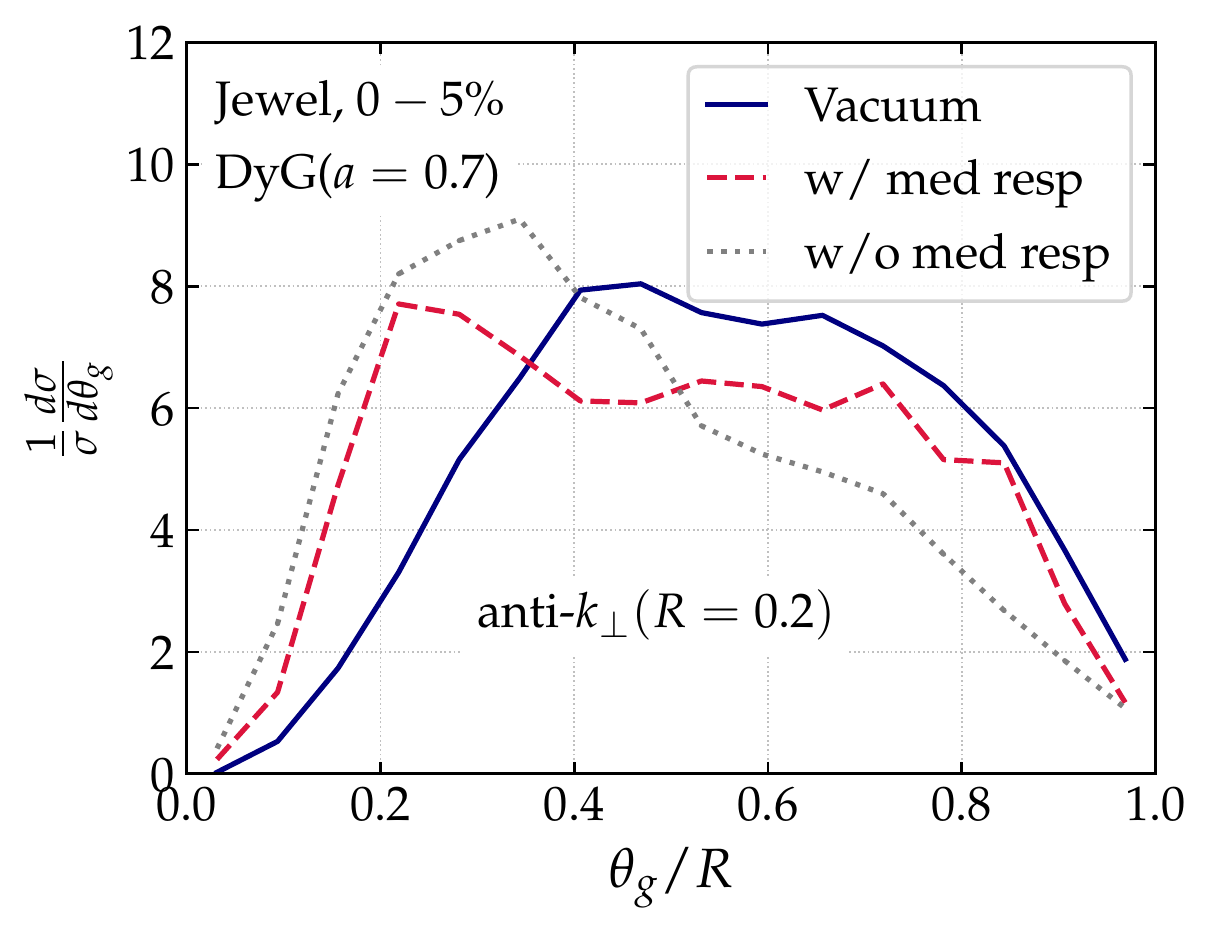}\includegraphics[scale=0.7]{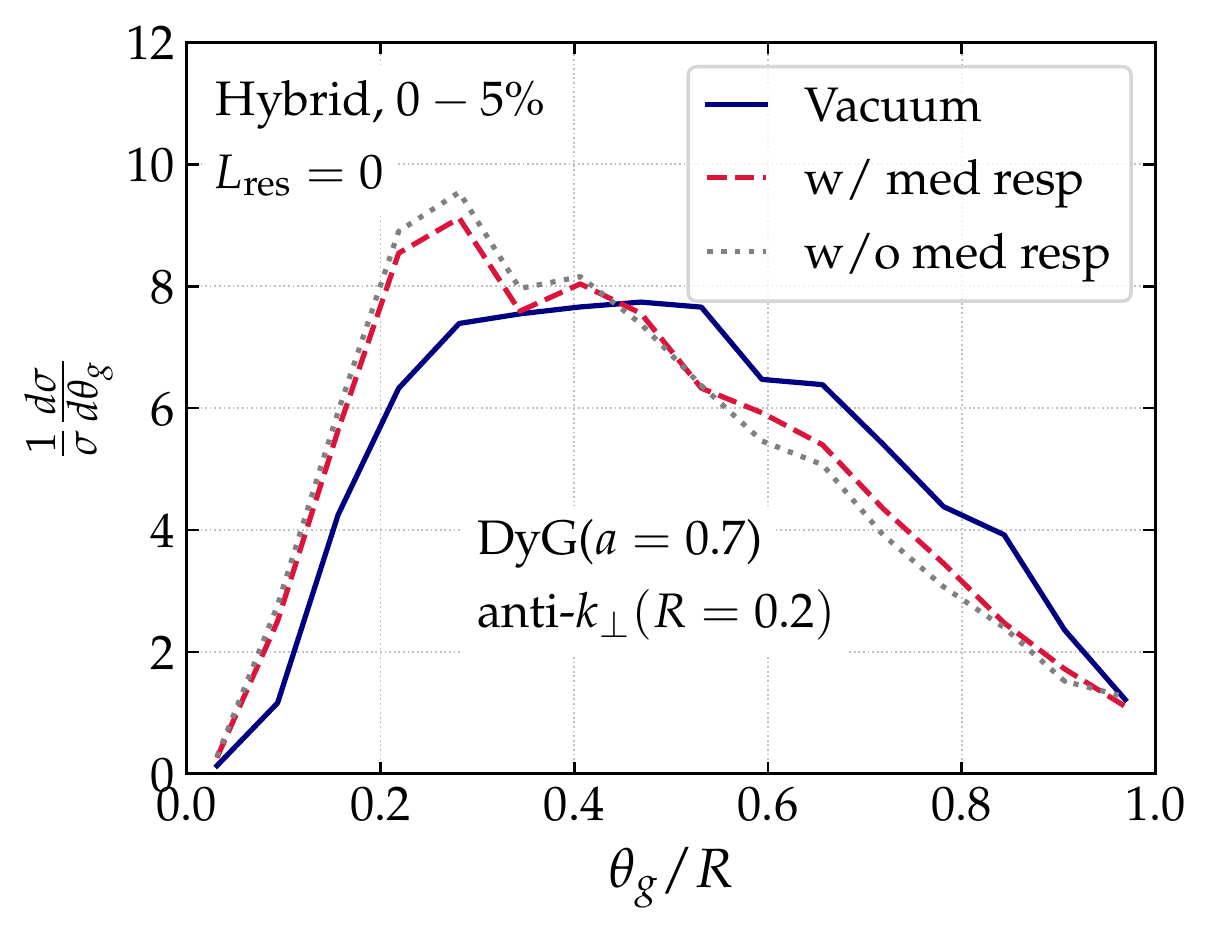}
    \caption{Same as Fig.~\ref{fig:med-response} but with $R=0.2$ showing a significant reduction of wide angle emissions originated from medium response.}
    \label{fig:med-response-rdep}
\end{figure*}
Several experimental measurements of jet substructure~\cite{Acharya:2018uvf,Acharya:2019djg,ATLAS-CONF-2019-056,ALICE:2021obz} have used small-$R$ jets aiming at reducing the impact of combinatorial jets. From the theoretical point of view, describing the cone-size dependence of jet quenching is an active field of research~\cite{Pablos:2019ngg,Mehtar-Tani:2021fud}. The impact of reducing the jet radius from $R=0.4$ to $R=0.2$ on the $\theta_g$-distribution is shown in Fig.~\ref{fig:med-response-rdep}. Clearly, tagging soft, thermal particles in narrower jets in less probable than in Fig.~\ref{fig:med-response}. In the case of the {\tt{Hybrid}} model this choice is extremely efficient in minimising the influence of medium response. The last statement is true for all values of the dynamical grooming parameter $a$. On the other hand, reducing the jet radius is not enough to make the sensitivity to recoil particles in {\tt{Jewel}} to vanish. Probably, an extra cut on the $z$ of the emissions \'a la SoftDrop would be helpful. Of course, by narrowing the phase-space for emissions, quenching effects are also diminished and that is the reason why the vacuum and medium distributions look more alike than in the $R=0.4$ case. Therefore, we conclude that, as expected, mitigating the impact of medium response by shrinking the jet radius comes at the price of a reduction in the discriminating power of the observable. 
\subsubsection{Centrality dependence}
\begin{figure}
   \includegraphics[width=\columnwidth]{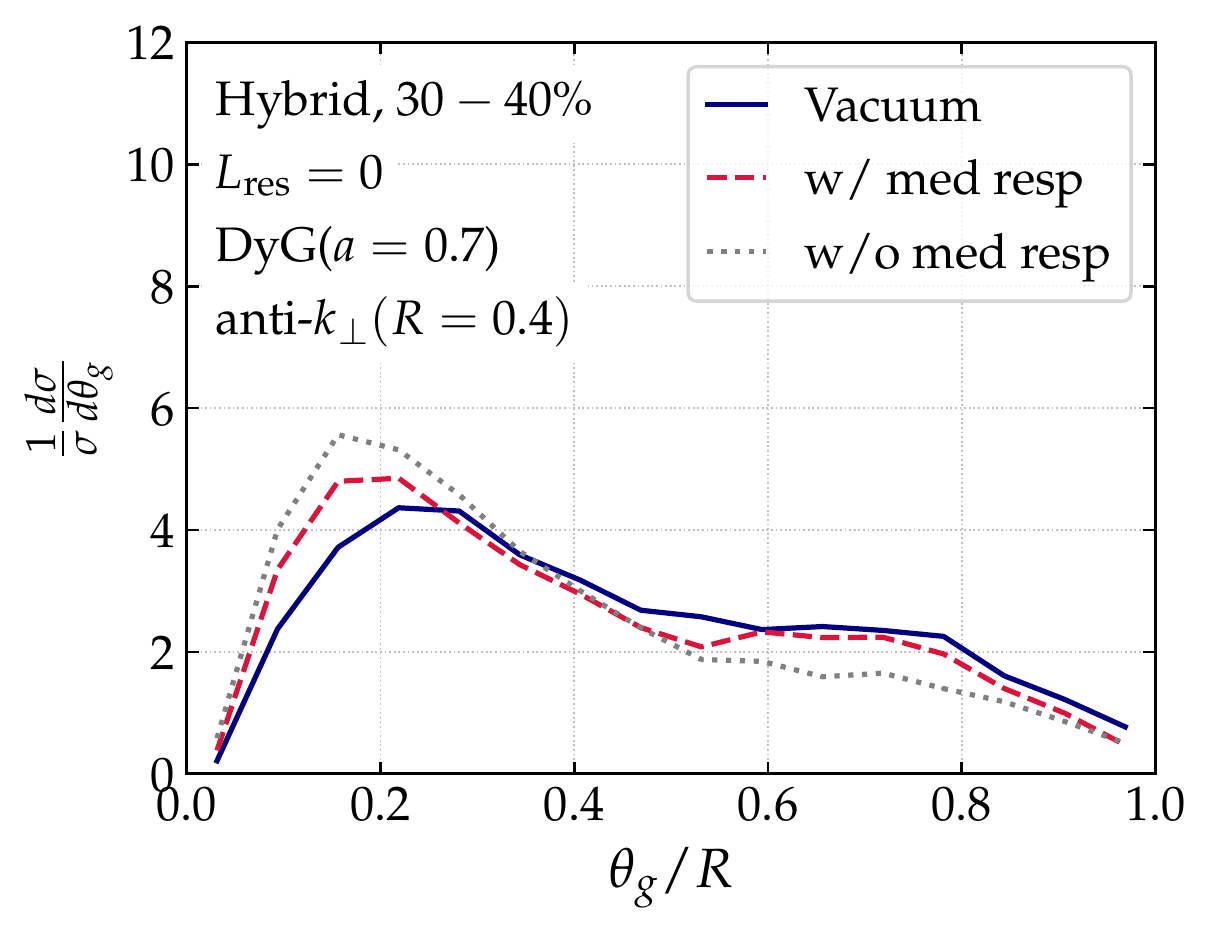}
    \caption{Same as Fig.~\ref{fig:med-response} but for $30-40\%$ in the {\tt{Hybrid}}. The medium response contribution is reduced but less effectively than in Fig.~\ref{fig:med-response-rdep}.}
   \label{fig:med-response-centdep}
\end{figure}
Another possibility to reduce the impact of medium response is to explore semi-peripheral collisions where the medium is not as dense as when the two nuclei collide head-on. Fig.~\ref{fig:med-response-centdep} demonstrates that moving to semi-peripheral collisions does not reduce the medium response component as effectively as reducing the jet radius did.  We therefore conclude that the combination that maximises the sensitivity to color coherence effects is: $R=0.2$, $0-5\%$ and $0.5\leq a \leq 1$.

Studying the dependence of the $\theta_g$-distribution with centrality is interesting not only from the point of view of reducing the impact of medium response, but also to further constraint the resolution angle~\cite{Mehtar-Tani:2021fud}. In a weak couplingg picture the scaling of $\theta_c$ with respect to the length of the medium is well known to be $\theta_c\propto L^{-3/2}$. In contrast, if no coherent angle existed and the maximum of the $\theta_g$-distribution was driven by the filtering effect caused by the finite size of the medium, one would expect a $\theta_{\rm max}\propto L^{-1}$ scaling. Therefore, the centrality dependence of the $\theta_g$-distribution's peak would be more strong in case energy loss was driven by $\theta_c$. In order to explicitly demonstrate this statement one would have to implement the geometry of the collision in {\tt{JetMed}} in such a way that different centralities can be simulated. This task is left for future work.  
%
\section{Final remarks and outlook}
\label{sec:conclusions}
In this paper, we present a comprehensive analysis of a promising jet substructure observable in heavy-ion collisions: the dynamically groomed jet radius $\theta_g$ distribution. We have shown through analytic calculations grounded in pQCD that the medium-modified distribution is strongly sensitive to the coherence angle of the medium $\theta_c$. These analytic calculations are supplemented by Monte-Carlo calculations of this observable with {\tt{JetMed}} which are in qualitative agreement with our analytic results. In summary, our study demonstrates the ability of the dynamically groomed jet radius to measure decoherence effects in the substructure of a jet traveling through a dense QCD medium. The numerical routines used in this work are provided in~\cite{python_git}. 

Given the simplifications inherent to our analytic and {\tt{JetMed}} calculations, it is not our intention to provide quantitative predictions for the $\theta_g$ distribution to be measured in heavy-ion collisions. However, we have been able to pin down the dominant physical mechanisms at stake which drive the modification of this observable. In fact, there are several dynamical process in the medium that converge at the QGP resolution angle and leave their imprint into the dynamically groomed jet radius. For instance, (i) the presence of the veto region leads to a narrowing of the $\theta_g$-distribution that is more pronounced, the larger the value of $a\le 2$ is. On the contrary, (ii) the medium induced branching kernel generates an enhancement of large-angle splittings due to transverse momentum broadening. The last two competing physical ingredients are eclipsed by (iii) differential energy loss. That is, when constructing a toy model for an in-medium parton shower that includes vacuum-like and medium induced emissions as well as energy loss, we observe that narrow splittings are enhanced for all values of $a$ with respect to the vacuum baseline. Notably, the inclusion of a fluctuating jet path length smoothens out the transition at $\theta_c$, but does not wash out the signal completely. 

We have also explored the sensitivity of our results to the underlying theoretical modelling of the jet-medium interactions, using the {\tt{Hybrid}} model that relies on strong coupling jet-medium interactions. This model also predicts a strong sensitivity of the observable to the medium resolution length, which is the strong coupling analogous of the coherence angle. Further, we have studied the impact of medium response in this observable with both the {\tt{Hybrid}} and {\tt{Jewel}} models. The imprint of these soft particles on the $\theta_g$-distribution is an enhancement of wide angle splittings. However, the magnitude of the $\theta_g\approx R_{\rm jet}$ peak significantly differs in the two descriptions of the medium. Since we are interested in a pQCD dominated observable we explore two routes to reduce the medium response contribution: reducing the jet cone size and using semi-peripheral events. Our findings indicate that the former option is more efficient than the latter.    

We would like to emphasise that the main difference between the in-medium and vacuum $\theta_g$ distributions it's not just a displacement in the peak position, but rather a significant modification of the shape of the distribution as a whole. That's the main reason why we quantify the in-medium to vacuum differences with the Kolmogorov-Smirnov distance and not with a simple ratio as typically done experimentally for other jet substructure observables such as the SoftDrop family in Ref.~\cite{ALICE:2021obz}. Thanks to the use of this metric, we are able to provide, both for the analytic calculation and for the Monte-Carlo simulations, a reasonable window for the Dynamical Grooming parameter $a$ and jet radius $R$ to be used experimentally in order to maximise the effects of the coherence angle, while at the same time minimising the non-perturbative contributions, such as hadronisation, geometry fluctuations and medium response. This is a tremendous advantage of the Dynamical Grooming procedure which depends on a single free parameter $a$. As such, studying the physics probed by the observable as a function of this single parameter is straightforward. We find that the optimal values are $0.5\lesssim a\lesssim 1$ and $R\lesssim 0.2$.

We have shown that both weak and strong couplingg models lead to similar trends in the $\theta_g$-distribution. In order to move forward and disentangle between theoretical models of jet quenching, there is a crucial need for performing global analyses in which models are tested against both global jet energy loss ($R_{AA}$ like) and jet substructure measurements, such as $\theta_g$ after Dynamical Grooming considered in this paper. A scan in terms of centrality classes or colliding system sizes is an interesting possibility to be explored in the future, given the theoretically well-defined path-length dependence of the critical angle. It will likely constrain more precisely the shape of the medium-modified phase space in Fig.\,\ref{fig:medium-lp} and reveal unambiguously the existence of a critical line at $\theta=\theta_c$ as well as its dependence on the physical properties of the medium. In addition, experimental data on the $k_t$ of the hardest emission would provide complementary information to the $\theta_g$ measurement since it probes the orthogonal direction in phase-space.

A natural extension of this work would be to go beyond the multiple soft scattering approximation of the parton-medium interaction. We plan to study the impact of rare, hard scatterings on the phase-space of emissions within the improved opacity expansion in a forthcoming publication~\cite{kt:paper}. Further, our resummation could be extended to account for heavy quarks in order to quantify the potential of the $\theta_g$-distribution to expose the dead-cone effect in heavy-ion collisions~\cite{deadcone:paper,Cunqueiro:2018jbh,ALICE:2021aqk,Dokshitzer:1991fd,Armesto:2003jh}. 
\section*{Acknowledgements} 
We are grateful to Daniel Pablos and Raghav Kunnawalkam Elayavalli for providing the Hybrid and Jewel samples, respectively. Enlightening conversations with Yacine Mehtar-Tani and Konrad Tywoniuk are acknowledged. We thank Liliana Apolinario, Leticia Cunqueiro, Laura Havener, Edmond Iancu, Yacine Mehtar-Tani, James Mulligan and Daniel Pablos for a careful reading of the manuscript and useful comments. P.C's work was supported by the U.S. Department of Energy, Office of Science, Office of Nuclear Physics, under contract No. DE- SC0012704. A.S.O.’s work was supported by the European Research Council (ERC) under the European Union’s Horizon 2020 research and innovation programme (grant agreement No. 788223, PanScales). A.T. work was supported by the Starting Grant from Trond Mohn Foundation (BFS2018REK01).

\appendix
\section{Sudakov with veto region}
\label{app:sudveto}
The purpose of this appendix is to provide a semi-analytic formula for the Sudakov form factor that includes the veto region for vacuum-like emissions given by Eq.~\eqref{eq:qw}. Using Eq.~\eqref{eq:veto-cond}, one can express $\Delta_{\notin\rm veto}$ in terms of $\Delta_{\in \rm veto}$ using
\begin{equation}
\ln(\Delta_{\notin \rm veto})=\ln(\Delta)-\ln(\Delta_{\in \rm veto})
\end{equation}
After replacing $\widetilde{P}(z,\theta)$ by Eq.~\eqref{eq:branch-dla}, we get for the in-veto contribution:
\beq
\ln\Delta_{\in\rm veto}(\kappa|a)  &= -2\bar\alpha\displaystyle\int_{\theta_{\rm min}}^{\theta_{\rm max}} \dis\frac{\dd \theta'}{\theta'}
\ln\dis\frac{z_{\rm max}(\kappa,\theta')}{z_{\rm min}(\kappa,\theta')}
\eeq 
with 
\begin{align}
	&z_{\rm min}(\kappa,\theta') = {\rm max}\left[\kappa\left(\frac{R}{\theta'}\right)^{a}, \frac{2}{\theta'^2p_tL}\right]\, ,  \\
	&z_{\rm max}(\kappa,\theta') ={\rm min}\left[1,\left(\frac{2\hat q}{p_t^3\theta'^4}\right)^{\frac13}\right]\, , \\
	&\theta_{\rm min}(\kappa) =  {\rm max}\left[R\kappa^{\frac1a},\theta_c,{\rm if } (a>\tfrac43): \left(\frac{2\hat q}{(\kappa R^ap_t)^3}\right)^{\frac{1}{4-3a}} \right] \, ,\\
	&\theta_{\rm max}(\kappa) =  {\rm min}\left[R, {\rm if } (a\leq\tfrac43): \left(\frac{2\hat q}{(\kappa R^ap_t)^3}\right)^{\frac{1}{4-3a}}\right]\, .
\end{align}
\section{Realistic jet spectrum}
\label{app:spectrum}
\begin{table}[hbt]
	\centering\small
	\begin{tabular}{|c|c|c|c|c|c|c|}
		\hline
		 & $i$ & $a$ [mb] & $p_{t0}$ [GeV] & $b$ &  $c$ \\
		\hline
		\multirow{2}{*}{{\bf PDF}} & $q$ & $0.008$ & $20.8$ &$4.7$ & $0.08$ \\ 
		\cline{2-6} 
		& $g$  & $0.007$ & $26.4$ &$5.1$ & $0.17$ \\
		\hline
		\multirow{2}{*}{{\bf nPDF}} & $q$ & $0.001$ & $32.8$ &$4.7$ & $0.1$ \\ 
		\cline{2-6} 
		& $g$ & $0.0005$ & $44.4$ &$5.2$ & $0.15$  \\
		\hline
	\end{tabular}
	\caption{\label{table:spectrum} The parametrization of the jet spectrum for both proton PDFs and using nuclear PDF for Pb.}
\end{table}
\begin{figure}[hbt]
     \includegraphics[width=\columnwidth]{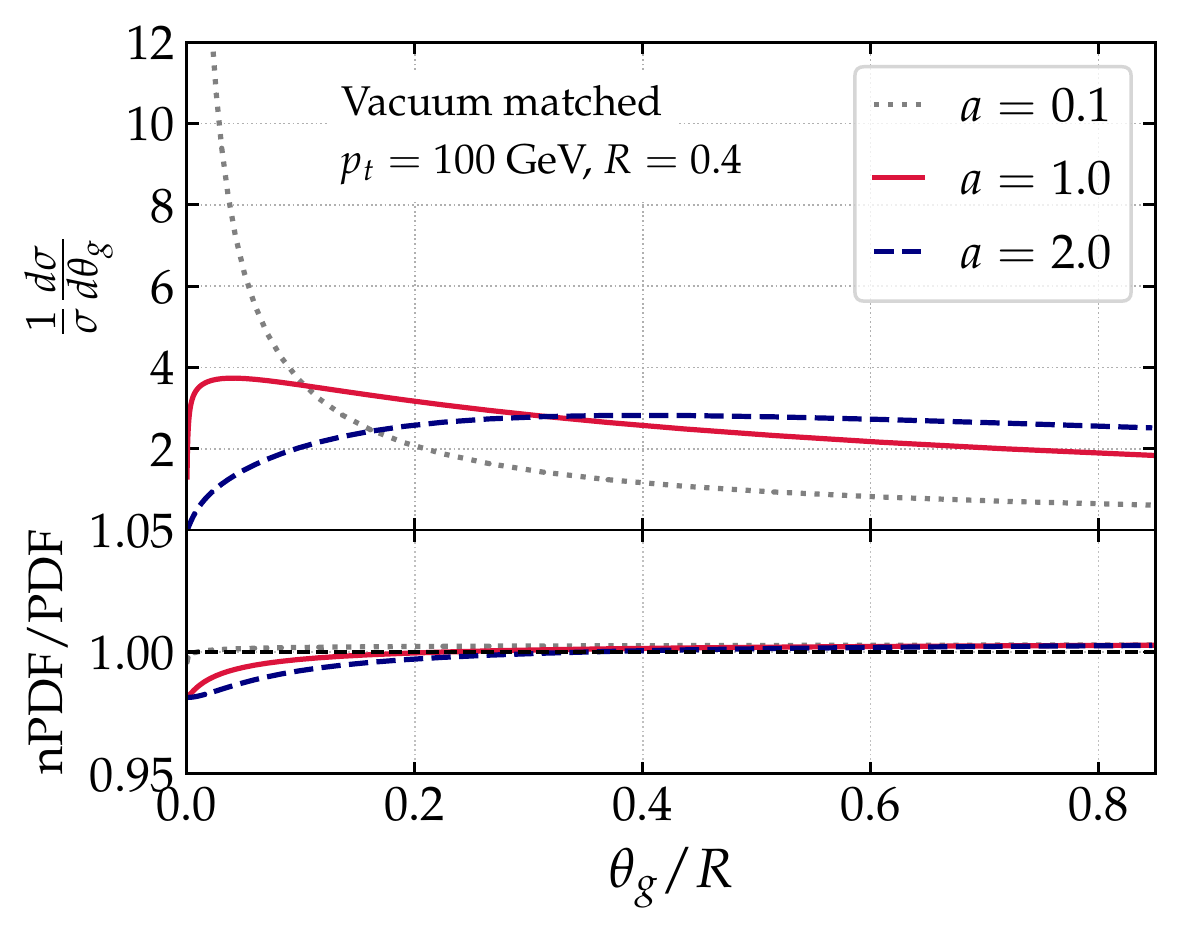}
    \caption{The $\theta_g$-distribution for vacuum splittings with realistic quark-gluon fractions using nuclear PDFs. The ratio to vacuum PDF is presented in the bottom panel showing a very mild impact of the nPDFs.}
    \label{fig:npdf}
\end{figure}
To make our analytic predictions closer to reality we have to include the jet spectrum. It enters in our calculation through the quark/gluon ratio and through the spectrum power $n$ in the energy loss component of Eq.~\eqref{eq:qw-zero}. We use the dijet parametrization from Ref.~\cite{Takacs:2021bpv} at $\sqrt{s}=5.02$~TeV, $|\eta|<2.8$ and $R=0.4$: 
\begin{align}
		\frac{\dd \sigma_i}{\dd p_{t}} &= a\left(\frac{p_{t0}}{p_t}\right)^{-(b+c\ln\frac{p_t}{p_{t0}})}\,,
\end{align}
where the subscript $i$ indicates the flavor of the initiating parton and $(a,b,p_{t0},c)$ are free parameters. This leads to the coefficients presented in Table~\ref{table:spectrum}.

The impact of the nuclear PDFs in the observable at play is shown in Fig.~\ref{fig:npdf}. We observe that the quark/gluon fraction is barely modified in this $p_t$ window.
\newpage
\bibliographystyle{apsrev4-1}
\bibliography{dyg-hic.bib}
\end{document}